\documentclass[review]{elsarticle}
\pdfoutput=1

\usepackage{hyperref}

\usepackage{amsmath}
\usepackage{amssymb}
\DeclareMathOperator*{\argmin}{argmin}
\usepackage{booktabs}
\usepackage{multirow}
\usepackage{siunitx}
\usepackage{subcaption}

\usepackage{rotating}

\usepackage{xcolor}
\definecolor{v2}{RGB}{31, 119, 180}

\newcommand{\sdot}{\boldsymbol{\cdot}} 

\journal{Journal of Computational Physics}

\begin{document}

\begin{frontmatter}

\title{Gradient Information and Regularization for Gene Expression Programming to Develop Data-Driven Physics Closure Models}

\author[melbourne]{Fabian Waschkowski\corref{correspondingauthor}}
\cortext[correspondingauthor]{Corresponding author}
\ead{fwaschkowski@student.unimelb.edu.au}

\author[beijing]{Haochen Li}
\author[aachen]{Abhishek Deshmukh}
\author[aachen]{Temistocle Grenga}
\author[beijing]{Yaomin Zhao}
\author[aachen]{Heinz Pitsch}
\author[melbourne]{Joseph Klewicki}
\author[melbourne]{Richard D. Sandberg}

\address[melbourne]{Department of Mechanical Engineering, University of Melbourne, VIC 3010, Australia}
\address[beijing]{Center for Applied Physics and Technology, Peking University, Beijing 100871, China}
\address[aachen]{Institute for Combustion Technology, RWTH Aachen University, 52062 Aachen, Germany}

\begin{abstract}

Learning accurate numerical constants when developing algebraic models is a known challenge for evolutionary algorithms, such as Gene Expression Programming (GEP). This paper introduces the concept of adaptive symbols to the GEP framework by Weatheritt and Sandberg (2016) \cite{weatheritt2016} to develop advanced physics closure models. Adaptive symbols utilize gradient information to learn locally optimal numerical constants during model training, for which we investigate two types of nonlinear optimization algorithms. The second contribution of this work is implementing two regularization techniques to incentivize the development of implementable and interpretable closure models. We apply $L_2$ regularization to ensure small magnitude numerical constants and devise a novel complexity metric that supports the development of low complexity models via custom symbol complexities and multi-objective optimization. This extended framework is employed to four use cases, namely rediscovering Sutherland's viscosity law, developing laminar flame speed combustion models and training two types of fluid dynamics turbulence models. The model prediction accuracy and the convergence speed of training are improved significantly across all of the more and less complex use cases, respectively. The two regularization methods are essential for developing implementable closure models and we demonstrate that the developed turbulence models substantially improve simulations over state-of-the-art models.

\end{abstract}


\end{frontmatter}

\section{Introduction}
\label{sec:introduction}

The development of physics models is at the core of most science and engineering disciplines. With the goal of understanding physical systems, predicting their future states and discovering governing laws, researchers have been developing physics models for centuries. The ability to predict the dynamics of a system-of-interest allows the design of advanced technologies, from microprocessors to spacecraft engines. Traditionally, such predictive models have been conceived based on physical, mathematical and empirical insights \cite{durbin2018}. Due to high costs of physical and numerical experiments in the past, data was limited and primarily used to verify and calibrate models \cite{montans2019}.

The rapid increase in computational resources in the last decades, however, enabled the generation of large quantities of high-fidelity data. Additional advances of machine learning (ML) algorithms, which are able to utilize big datasets, motivated a paradigm shift from traditional model development to data-driven modeling in various scientific fields \cite{montans2019}. The most frequently applied ML models are deep neural networks (DNN) \cite{goodfellow2016}, which demonstrated remarkable success on previously intractable problems, from computer vision initially to protein folding more recently \cite{krizhevsky2012,jumper2021}. While DNNs posses excellent predictive capabilities on high-dimensional datasets, the trained models are difficult to interpret and have been shown to generalize poorly to data outside of the training distribution for nonlinear target functions \cite{xu2021a}. Furthermore, the models resulting from DNN training or decision tree-based methods, another often applied type of ML algorithm, are highly complex equations or algorithmic models. For the application as closure models, i.e. the implementation in underdetermined systems of equations, which are the focus of this work, these models are difficult to use due to stability issues during the numerical solution of equation systems.

An alternative to such highly complex models are algebraic models, which are typically the result of traditional model development. However, \citet{schmidt2009} used genetic programming (GP) \cite{koza1992}, which is an evolutionary algorithm, to symbolically regress algebraic models from experimental data. This data-driven approach yields interpretable and implementable models. \citet{cranmer2020} showed that algebraic models derived via symbolic regression even generalize better to out-of-distribution data than DNNs for different physical problems. 

\citet{ferreira2001} developed with gene expression programming (GEP) an algorithm that improves over GP by introducing a genotype-phenotype distinction (see Section \ref{sec:standard} for details), as inspired by natural evolution. While symbolic regression via GP or GEP is popular due to a high flexibility in the resulting model structure, the problem of not converging to accurate numerical constants in the model equation is well known \cite{ryan2003,zhong2017}. Typically, a finite number of numerical constants are created before the training and other constants in the model equation are only achieved by combining existing constants via mathematical operators. Thus, converging to specific constants is challenging, especially since evolutionary algorithms are fundamentally stochastic. 

A different approach for the development of algebraic models was devised by \citet{brunton2016}, which performs linear regression on a library of equation snippets with sparsity-enforcing regularization. This approach, which is termed sparse regression, computes numerical constants accurately using a deterministic linear least squares solver. A considerable downside is, however, that the possible model structures are limited to linear combinations of the equation snippets in the library.

The goal of this paper is combining the advantages of symbolic and sparse regression to develop data-driven closure models with flexible structures and accurate numerical constants. We extend the GEP framework developed by \citet{weatheritt2016} to incorporate gradient information to optimize model constants. In the literature, both gradient-free and gradient-based methods for constant optimization in GEP exist. \citet{li2004} investigated random and creep mutation operators that modify numerical constants during training. Other gradient-free methods applied to this problem are hill climbing \cite{lopes2004} and differential evolution \cite{zhang2007}. To improve over these stochastic methods, \citet{zarnegar2009} employed a linear least squares solver to calculate the values of constants pre-multiplied to trained basis functions. \citet{dominique2021} defined a special power operator, of which the exponent was determined using a gradient-based nonlinear least squares solver.

A novelty of this paper is the introduction of so-called adaptive symbols, which allow gradient-informed numerical constants in GEP in a general form, i.e. at any position in the trained model equation. To determine the values of adaptive symbols $p$ during training, both a general optimizer to solve

\begin{equation}
    p^* = \argmin_{p \in \mathbb{R}^n} J(p) \, , \quad \lVert p - p^* \rVert < \delta \, , \quad \delta > 0
    \label{eq:minimizer}
\end{equation}

\noindent for any objective function $J \colon \mathbb{R}^n \to \mathbb{R}$ and a nonlinear least squares optimizer to solve Eq. \eqref{eq:minimizer} for 

\begin{equation}
    J(p) = \frac{1}{2} \sum_{i=1}^m r_i(p)^2 \, , \quad r \colon \mathbb{R}^n \to \mathbb{R}^m \, , \quad m \ge n
    \label{eq:leastsquares}
\end{equation}

\noindent are investigated \cite{nocedal1999}. Furthermore, we apply two regularization methods to avoid overfitting the training data and incentivize the development of implementable models. Small magnitudes of numerical constants and a low complexity of the evolved model equation are considered indicators for the numerical stability of closure models. Thus, we implement $L_2$ regularization \cite{goodfellow2016} and define a model complexity metric, which is set as an additional objective function similar to the approach for GP of \citet{schmidt2009}.

This new gradient-informed and regularized GEP framework is applied to four different use cases. First, we rediscover Sutherland's viscosity law from generic data as a proof-of-concept. Next, laminar flame speed models are trained on data of unstretched premixed flames. Finally, we develop two types of models for turbulent flow. A subgrid-scale (SGS) model for large eddy simulations (LES) of homogenous isotropic turbulence and a nonlinear eddy viscosity model (NLEVM), including a turbulence production correction model, for Reynolds-averaged Navier-Stokes (RANS) calculations of a three-dimensional flow around a wall-mounted square cylinder are trained.

The structure of this paper is as follows. We briefly discuss the standard GEP framework in Section \ref{sec:standard} and then introduce the concept of adaptive symbols and the optimization algorithms to determine their values in Section \ref{sec:adaptivesymbols}. In Section \ref{sec:regularization}, the regularization techniques that support generalizability and implementability of the developed models are described. Section \ref{sec:results} is organized such that each use case is introduced in detail with its modeling and training strategies before the respective training results are analyzed. Conclusions are drawn in Section \ref{sec:conclusion}.

\section{Methodology}
\label{sec:methodology}

The novel contributions of this paper are the introduction of adaptive symbols and the application of $L_2$ regularization and a model complexity objective function to GEP. The former is discussed in Section \ref{sec:adaptivesymbols}, while the latter is described in Section \ref{sec:regularization}. We start by outlining the utilized GEP framework.

\subsection{Standard GEP framework}
\label{sec:standard}

\begin{figure}[!ht]
	\centering
    \resizebox{0.75\linewidth}{!}{
		\centering
        \includegraphics{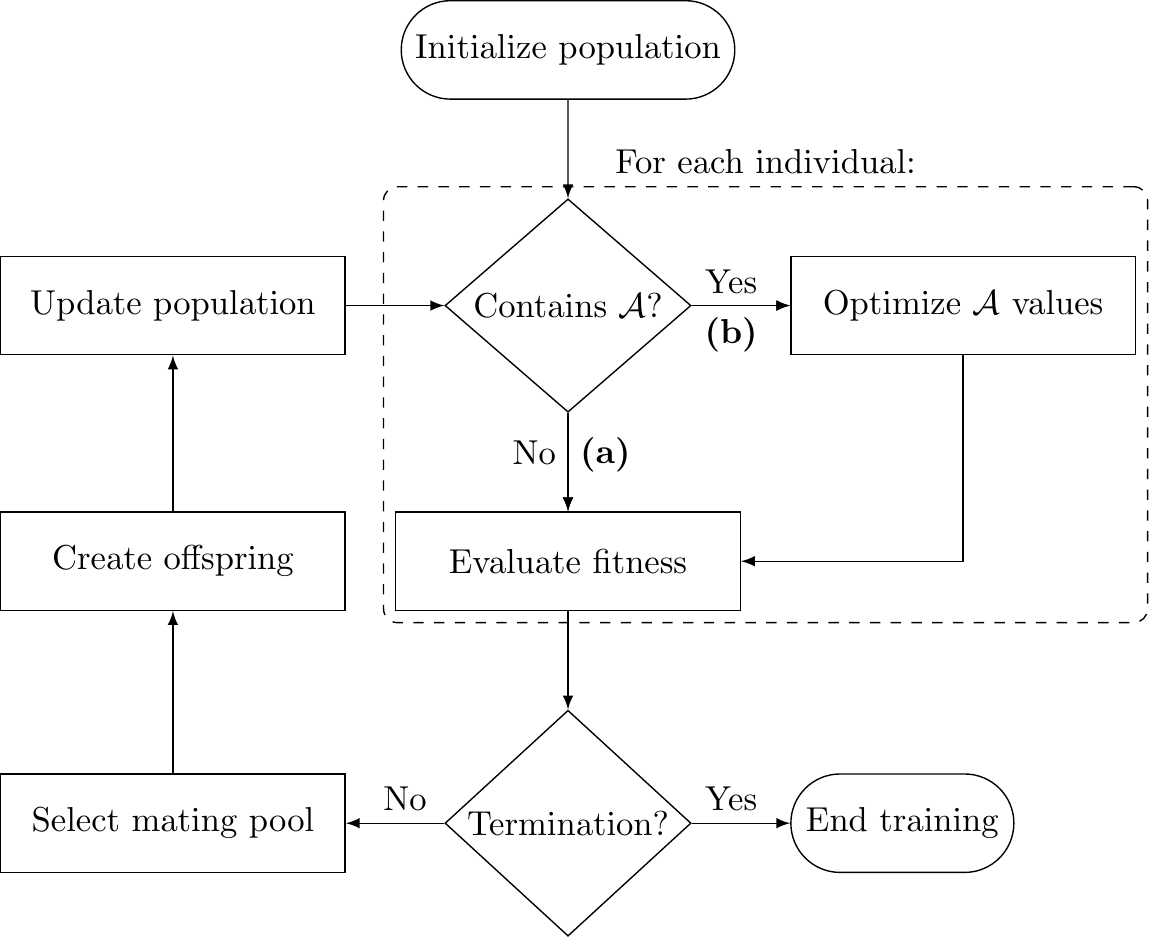}
	}
	\caption{Flowchart of the GEP framework without (a) and with (b) the extension to handle adaptive symbols ($\mathcal{A}$).}
	\label{fig:flowchart}
\end{figure}

\citet{weatheritt2016} developed the GEP framework employed in this paper to enable tensor regression, i.a. for turbulence closure modeling. The framework implements the GEP algorithm by \citet{ferreira2001}, which is a type of evolutionary algorithm. As such, a population of candidate models, so-called individuals, is evolved over numerous generations to minimize a specific training objective. Fig. \ref{fig:flowchart} (a) illustrates the corresponding flowchart. Initially, a population of random individuals is created. In each generation, the fitness of each individual according to the training objective is evaluated. If the set termination criterion is not fulfilled, individuals compete based on their fitness for selection to the mating pool. Possible termination criteria are a minimum fitness in the population, a specified number of generations or a maximum training runtime. Next, offspring are created by applying genetic operators, such as mutation or crossover, to individuals in the mating pool. Finally, the offspring update the population by replacing unfit individuals.

\begin{figure}[!ht]
	\centering
    \resizebox{.95\linewidth}{!}{
        \centering
        \includegraphics{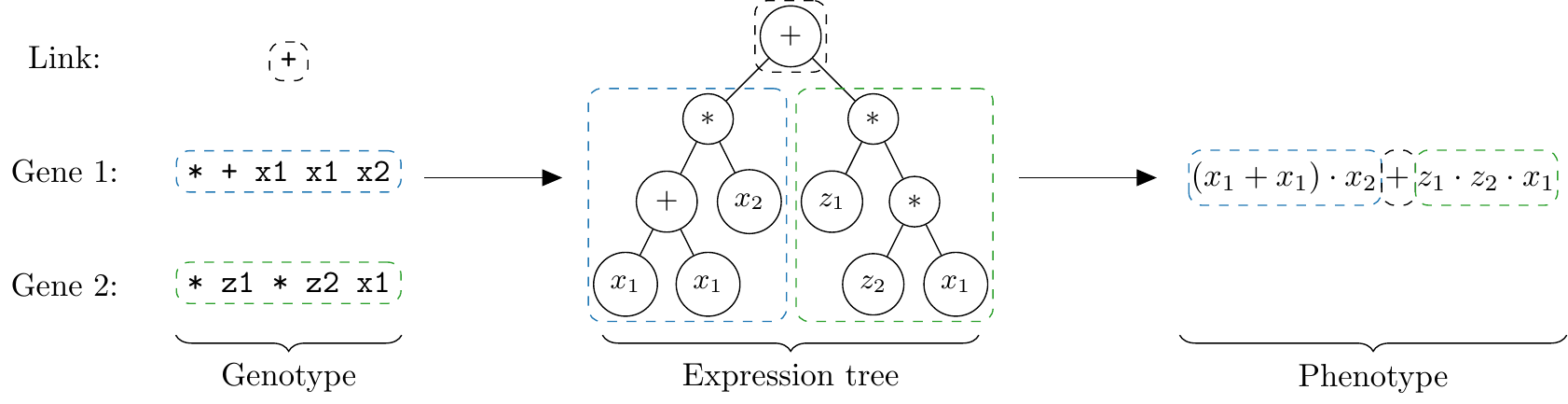}
	}
	\caption{Translation from genotype to phenotype of exemplary individual consisting of two genes (blue, green) and one link (black).}
	\label{fig:translation}
\end{figure}

Structurally, individuals in GEP consist of multiple genes and each gene is encoded as a linear string of symbols, which is referred to as its genotype. Typical symbols are input variables, mathematical operators and numerical constants. To evaluate the fitness of an individual, the genotype of each gene is translated to a nonlinear expression tree, which can be interpreted as an algebraic equation. This equation is referred to as the gene's phenotype. Then, these phenotypes are linked to yield the complete algebraic model, i.e. the individual's phenotype, and the training objective is calculated. Fig. \ref{fig:translation} shows the translation of an individual with two genes for the input variables $x_1$ and $x_2$, addition and multiplication operators and the numerical constants $z_1$ and $z_2$.

In comparison to genetic algorithms (GA) \cite{holland1975} and GP, the advantage of GEP results from the distinction between the genotype and the phenotype. While genetic operators are applied to the genotype, the phenotype of an individual determines its fitness. This distinction allows easy genetic manipulations and complex functional expressions. In contrast, genes in GA are implemented as linear strings only, which limits their functional expressivity, and genes in GP are represented as expression trees only, which complicates genetic manipulation.

\subsection{Adaptive symbols and numerical optimizers}
\label{sec:adaptivesymbols}

The concept of adaptive symbols is introduced in GEP to generate algebraic models with accurate numerical constants and to determine these constants efficiently, i.e. based on gradient information instead of stochastic processes. Adaptive symbols are devised as a new type of symbol that are, in addition to standard symbols like input variables, mathematical operators or fixed numerical constants, available to the GEP algorithm to build and mutate the genotype of individuals in the population.

In the genotype, adaptive symbols behave precisely like standard symbols, so that the same genetic operators can be applied. In the phenotype, however, adaptive symbols act as placeholders. When one or more of these placeholders are detected prior to an individual's fitness evaluation, a process is started to determine the optimal numerical constants at the placeholder positions (see Fig. \ref{fig:flowchart} (b)). In this process, an iterative gradient-based optimizer is employed with the goal of minimizing the specified training objective. The resulting constants, i.e. the adaptive symbol values, are inserted at the placeholder positions and the fitness of the individual is calculated. Finally, each adaptive symbol stores its value as an initial value for next generation's fitness evaluation. The following aspects further define the concept of adaptive symbols:

\begin{enumerate}
    \item Adaptive symbols can occur at any position in the phenotype.
    \item Values of adaptive symbols are unique for each individual, not across the population.
    \item The number of adaptive symbols per individual is user-defined.
    \item One adaptive symbol can be selected to the genotype multiple times.
    \item All instances of one adaptive symbol share the same value in the phenotype.
\end{enumerate}

As an example, data generated from the canonical function $f(x_1, x_2) = 0.196 \, x_1^2 + 0.616 \, x_2 + 3.142$ could be approximated using the input symbols $x_1$ and $x_2$, the operators $+$ and $*$ and two adaptive symbols $p_1$ and $p_2$ with the genotype

\begin{equation*}
    \texttt{+ * p1 * x1 x1 + * * p1 p2 x2 p2}
\end{equation*}

\noindent that translates to the phenotype 

\begin{equation*}
    p_1 \cdot x_1^2 + p_1 \cdot p_2 \cdot x_2 + p_2 \, ,
\end{equation*}

\noindent where the numerical optimizer would calculate $p_1 = 0.196$ and $p_2 = 3.142$.

\subsubsection*{Numerical optimizers}

Determining the locally optimal adaptive symbol values $p^*$ is generally a nonlinear optimization problem, as any training objective $J$ depends on the candidate model and adaptive symbols can occur at any position in the candidate model. Such a problem requires an iterative solution of Eq. \eqref{eq:minimizer} and we investigate two types of numerical optimizers.

First, a general optimizer is applied to solve Eq. \eqref{eq:minimizer} for any objective function $J \colon \mathbb{R}^n \to \mathbb{R}$, which provides a high flexibility for defining the training objective. Specifically, we select the Broyden-Fletcher-Goldfarb-Shanno (BFGS) algorithm \cite{broyden1970b,fletcher1970,goldfarb1970,shanno1970} based on the results of a preliminary study. The algorithm improves over the standard gradient descent algorithm by utilizing curvature information and is computationally efficient by approximating instead of calculating the inverse Hessian matrix of $J$. This approximation is updated at every iteration based on the observed change of gradients \cite{nocedal1999}.

The second type of investigated optimizers are nonlinear least squares optimizers, as most ML training objectives are formulated to minimize the square errors between the model predictions and the training data. Eq. \eqref{eq:leastsquares} defines such a training objective $J$, where the residual function $r$ describes the dependency of these prediction errors on the adaptive symbol values $p$. The mathematical structure of least squares problems can be exploited to approximate the Hessian matrix of $J$ based on the Jacobian matrix of $r$, which is considered to be an often accurate approximation \cite{nocedal1999}. The Levenberg-Marquardt (LM) algorithm \cite{levenberg1944,marquardt1963}, which is applied in the studies in Section \ref{sec:results}, combines this approximation of the Hessian with additional regularization to control the iteration step size. The BFGS and LM algorithms are implemented in the GEP framework via the SciPy package for Python \cite{virtanen2020}.

\subsection{Regularization and model complexity}
\label{sec:regularization}

Regularization is applied in ML in general to prevent a trained model from overfitting the training data. In other words, the model is incentivized to not memorize the training data points but approximate the underlying data generating function, so that the error on testing data from the same data distribution is close to the training error. For the development of closure models in particular, the experience of previous studies is that implementing complex models with large numerical constants to close underdetermined systems of equations can lead to instabilities when solving these systems numerically \cite{weatheritt2017a}. To control the magnitudes of adaptive symbol values and the complexity of the evolved algebraic models, $L_2$ regularization and a model complexity objective function are added to the GEP framework.

$L_2$ regularization is the most frequently used regularization technique and drives model parameters towards small magnitudes \cite{goodfellow2016}. Therefore, the square $L_2$ norm of the parameter vector $p \in \mathbb{R}^n$ is multiplied by a scalar regularization parameter $\lambda$ and added to the training objective $J$ to yield the extended training objective

\begin{equation}
    \hat{J} = J + \lambda \cdot \lVert p \rVert_2^2 = J + \lambda \cdot \sum_{i=1}^n p_i^2 \, ,
    \label{eq:l2regularization}
\end{equation}

\noindent where the number of parameters $n$ corresponds to the number of unique adaptive symbols in the phenotype of the evaluated individual. The regularization parameter $\lambda$ is user-defined and balances fitting the training data with reducing adaptive symbol magnitudes. Due to the nature of $L_2$ regularization, determining $p^*$ that minimizes $\hat{J}$ remains a least squares problem to which both investigated optimizers can be applied.

\subsubsection*{Model complexity}

The complexity of a model describes the capability of its model class $F$ to approximate a wide range of functions by adapting its parameters $p$ \cite{goodfellow2016}. One example of a model class are univariate polynomials of degree $d$

\begin{equation}
    F_d(x) = \sum_{i=0}^d p_i x^i \, .
\end{equation}

\noindent If we assume that exemplary training data is generated by a quadratic function, the complexity of any linear model in $F_1$ is too low to accurately fit the training data. On the other hand, the complexity of a polynomial of degree five is considered too high, as training a model in $F_5$ will likely overfit the training data. Thus, controlling the model complexity is another regularization technique to improve generalizability. Additionally, a low complexity naturally increases a model's interpretability and, as discussed above, is beneficial for the implementability of closure models.

Quantifying model complexity, however, is not a straightforward task, as no universally accepted metric exists. In GP, published metrics can be classified as calculating either the structural complexity or the functional complexity of a model. Structural complexity metrics analyze the genotype of an individual, which is an expression tree in GP, and measure, for example, the number of tree nodes or the sum of subtree nodes \cite{schmidt2009,smits2005}. In contrast, the functional complexity of a model depends on its phenotype, i.e. the resulting algebraic equation. \citet{vladislavleva2008} estimated the order of nonlinearity of models to measure complexity and \citet{vanneschi2010} devised a curvature-based metric. Furthermore, metrics from statistical learning theory were investigated, such as the Vapnik-Cervonenkis dimension or the Rademacher complexity \cite{chen2018,chen2020}. Complexity metrics in GP are either used to extend the objective function to a weighted composite function \cite{soule1998} or set as an additional objective function \cite{smits2005}.

In GEP, to the authors' knowledge, only \citet{ferreira2006a} studied model complexity and applied a metric based on the count of expression tree nodes. As structural complexity metrics are computationally inexpensive and successful at preventing ineffective symbols in the model equation, we also define a structural model complexity metric, but one that is based on the symbol complexity of expression tree nodes. Therefore, each symbol $s$ in the set of symbols $\mathcal{S}$ available to the GEP algorithm is assigned a user-defined complexity value $c_s$. The complexity of an evolved model is then calculated as 

\begin{equation}
    J_c = \sum_{s \in L} c_s \, ,
    \label{eq:complexity}
\end{equation}

\noindent where $L$ is the list of symbols in the individual's expression tree. The ability to define custom symbol complexities, in contrast to counting the number of symbols, which is equivalent to $c_s = 1 \;\, \forall \;\,  s \in \mathcal{S}$, allows incentivizing specific model structures. For example, setting $c_+ = 1$ and $c_{\exp} = 3$ for the symbols $s_+(x_1, x_2) = x_1 + x_2$ and $s_{\exp}(x) = \exp(x)$ supports the usage of linear over nonlinear operators. While \citet{ferreira2006a} extended the training objective via a weighted sum approach to account for model complexity, we apply Eq. \eqref{eq:complexity} as a second objective function. In comparison, our multi-objective approach does not require weighting the training error and the model complexity metric, which is difficult to estimate before the training, and enables the analysis of models of different complexity selected from the Pareto front (see Fig. \ref{fig:nlevmparetoA} and \ref{fig:nlevmparetoR}) after the training. We utilize the multi-objective optimization extension to the GEP framework by \citet{waschkowski2022}.

\section{Results}
\label{sec:results}

The application of the presented adaptive symbol concept (see Section \ref{sec:adaptivesymbols}) and regularization methods (see Section \ref{sec:regularization}) to four different use cases is discussed in the following. While all cases employ adaptive symbols, we investigate different optimizers and regularization techniques in each use case. 

First, we compare the performance of the BFGS and LM optimizers when rediscovering Sutherland's law in Section \ref{sec:sutherland}. Next, laminar flame speed models are developed for unstretched premixed flames and the advantages of the model complexity objective function $J_c$ are explored (see Section \ref{sec:combustion}). In Section \ref{sec:les}, the impact of $L_2$ regularization on the development of SGS models for LES of homogenous isotropic turbulence is demonstrated. Lastly, we combine the two regularization techniques and compare the two optimizers when training NLEVMs for RANS calculations of a wall-mounted square cylinder flow (see Section \ref{sec:rans}). Table \ref{tab:cases} presents an overview of the use cases.

\begin{table}[h]
	\centering
	\caption{Overview of studied use cases.}
	\begin{tabular}{l c c c}
        \toprule
        Use case & Optimizers & Regularization \\
        \midrule
        Sutherland's law & BFGS, LM & - \\
        Laminar flame speed modeling & BFGS & $J_c$ \\
        Subgrid-scale modeling & BFGS, LM & $L_2$ \\
        Nonlinear eddy viscosity modeling & BFGS, LM & $L_2$, $J_c$ \\
        \bottomrule
	\end{tabular}
	\label{tab:cases}
\end{table}

All results are compared to the standard GEP framework and a maximum training runtime is set as the termination criterion, unless otherwise stated. As optimizing the values of adaptive symbols increases the computational training costs, we aim to ensure an unbiased comparison between training runs with varying numbers of adaptive symbols by providing equal computational resources to all runs. In other words, training runs with no or few adaptive symbols are able to evolve their population for more generations, while runs with higher numbers of adaptive symbols can benefit from more gradient-informed numerical constants. In the following, the details of each case and its modeling and training strategies are presented before the respective results are discussed.

\subsection{Sutherland's law}
\label{sec:sutherland}

Sutherland's law \cite{sutherland1893} models the dynamic viscosity $\mu$ of dilute gases as 

\begin{equation}
    \mu = \mu_0 \left(\frac{T}{T_0} \right)^\frac{3}{2} \frac{T_0 + C}{T + C} \, ,
    \label{eq:sutherland}
\end{equation}

\noindent where $\mu$ is solely a function of the temperature $T$. The Sutherland temperature $C$ is a gas-specific constant and $\mu_0$ and $T_0$ are reference values. Sutherland's law is derived from kinetic gas theory and assumes an idealized intermolecular-force potential \cite{white2006}. In computational fluid dynamics (CFD) simulations of compressible flows, Eq. \eqref{eq:sutherland} is commonly applied as a closure model to describe the linear dependency of the viscous stresses of Newtonian fluids on the strain rate tensor.

We utilize Sutherland's law as a canonical example to discuss the benefits of adaptive symbols and analyze the two selected optimizers. We aim to rediscover first Eq. \eqref{eq:sutherland} and then its normalized version $\hat{\mu} = \mu / \mu_0$ for air with $C = \SI{110.4}{\kelvin}$ and $\mu_0 = \SI[per-mode = symbol]{1.716d-5}{\kilogram\per\metre\per\second}$ at $T_0 = \SI{273.15}{\kelvin}$. The training data for $\mu$ and $\hat{\mu}$ is generated according to Eq. \eqref{eq:sutherland} at 100 uniformly spaced data points in a range from $T = \SI{250}{\kelvin}$ to $T = \SI{1750}{\kelvin}$.

\subsubsection*{Modeling and training strategies}

Three strategies to rediscover Sutherland's law are investigated. We compare the standard training approach of using five random numerical constants (RNC) drawn from a uniform distribution $U(-1, 1)$ to using five adaptive symbols with values determined by either the BFGS or LM optimizer. 

The remaining symbols and hyperparameters are unchanged across the different training approaches. The single input symbol is $T$ and the mathematical operators are $+$, $-$, $\times$, $\div$ and $(\sdot)^\frac{3}{2}$. The integer values $0$, $1$ and $2$ are provided as additional numerical constants. In every training run, a population of 1000 individuals consisting of two genes each is evolved to minimize the mean squared error (MSE) between the predicted (normalized) viscosities and their training data values. For each of the three strategies and both $\mu$ and $\hat{\mu}$, we perform the training with five different random initializations of the population. The maximum runtime per training run is set to 0.25 CPU hours. An overview of all training settings is provided in Table \ref{tab:strategies}.

\subsubsection*{Analysis and discussion}

\begin{table}[ht]
	\centering
	\caption{Accuracy and training time (mean $\pm$ SD) for rediscovering Sutherland's law in its standard ($\mu$) and normalized form ($\hat{\mu}$) using the standard training strategy (RNC) or adaptive symbols with different optimizers (BFGS, LM).}
    \begin{tabular}{l c c c c}
        \toprule
        \multirow{2}{*}{Strategy} & \multicolumn{2}{c}{$\mu$} & \multicolumn{2}{c}{$\hat{\mu}$} \\
        \cmidrule(lr){2-3} \cmidrule(lr){4-5}
        & Acc. & Time [$\text{CPU} \, \si{\s}$] & Acc. & Time [$\text{CPU} \, \si{\s}$] \\
        \cmidrule(lr){1-5}
        RNC         & 0/5 & --  & 0/5 & --  \\
        BFGS        & 0/5 & --  & 4/5 & $355.7 \pm 209.4$ \\
        LM          & 5/5 & $217.4 \pm 185.0$ & 5/5 & $113.1 \pm 73.0$ \\            
        \bottomrule
    \end{tabular}
	\label{tab:SLresults}
\end{table}

The results of training models for $\mu$ and $\hat{\mu}$ to rediscover Sutherland's law using the three introduced training strategies are presented in Table \ref{tab:SLresults}. The accuracy describes the ratio of rediscoveries to randomly initialized training runs and the reported time is the mean training time for successful rediscoveries and its standard deviation (SD).

Training with five adaptive symbols and employing the LM optimizer allows to identify the correct $\mu$ and $\hat{\mu}$ equations in all training runs. In contrast, using the BFGS optimizer, the GEP framework is only able to learn Sutherland's law in its normalized form. 

Sutherland's law can be simplified to $\mu = C_1 T^\frac{3}{2} / (T + C)$, where the combined constant is $C_1 = \SI[per-mode=symbol]{1.458d-6}{\kilogram\per\metre\per\second\per\kelvin^{\frac{1}{2}}}$. For the normalized version $\hat{\mu}$, $C_1$ changes to $\SI[]{8.496d-2}{\per\kelvin^{\frac{1}{2}}}$. Thus, the difference in the order of magnitude between $C_1$ in the numerator and $C = \SI{110.4}{\kelvin}$ in the denominator is reduced significantly for $\hat{\mu}$, which enables rediscoveries using the BFGS optimizer and speeds up the training with the LM optimizer.

The standard training approach utilizing RNCs is not capable of rediscovering Sutherland's law in any training run, despite evolving populations for up to 798 generations (in contrast to a maximum of 114 generations with the LM optimizer). One issue that we identified is a conflict between model structure, constants and fitness. The fittest model resulting from the training with RNCs to rediscover $\hat{\mu}$ is 

\begin{equation}
    \hat{\mu} = \SI[]{1.469d-3}{\per\kelvin} \cdot T + 0.86
\end{equation}

\noindent with a fitness value of $J = 0.82 \times 10^{-2}$. This model has a structure and constants that are less similar to the simplified version of Sutherland's law than 

\begin{equation}
    \hat{\mu} = \SI[]{7.925d-2}{\per\kelvin\tothe{\frac{1}{2}}} \cdot T^\frac{3}{2} / T \, ,
\end{equation}

\noindent which is the least fit model resulting from the five training runs ($J = 2.1 \times 10^{-2}$).

Adaptive symbols are capable of resolving this conflict and demonstrate significant performance advantages on this canonical example. In particular, the least squares-specific LM optimizer robustly determines accurate numerical constants across varying orders of magnitude. While the results of the standard training approach could potentially be improved by modifying the RNC sampling distribution, adapting GEP hyperparameters or normalizing the training data differently, the concept of adaptive symbols simplifies the training procedure and adds flexibility to the GEP framework.

\subsection{Laminar flame speed modeling}
\label{sec:combustion}

The laminar flame speed $S_L$ is an important characteristic of a given fuel-air mixture in combustion engines. $S_L$ describes the rate of propagation of the flame surface in premixed combustion processes as a result of chemical reactions, mass diffusion and heat conduction. Numerical simulations of premixed combustion engines require an accurate description of the combustion processes and typically use empirical or analytical laminar flame speed models. The accuracy of these models affects the numerical prediction of the combustion behavior and the emission of pollutants significantly, which in turn influences the potential for improving the combustion processes. Therefore, obtaining accurate laminar flame speed models that apply to the wide range of operating conditions observed in industrial premixed combustion engines is a key challenge.

Among the most widely used $S_L$ models is the Gülder model \cite{gulder1984}, which is an empirical model fitted to experimental data. \citet{metghalchi1982} introduced an empirical power-law expression that accounts for the temperature and pressure dependency of $S_L$ and captures dilution effects. In contrast, \citet{gottgens1992} proposed a physics-based approach to derive an analytical expression for $S_L$ based on rate-ratio asymptotics. The constants in this analytical expression have been fitted to a wide range of fuels, such as hydrogen, methane, ethylene, ethane, acetylene, propane, ethanol, n-heptane, iso-octane and primary reference fuel \cite{gottgens1992, rohl2009, ewald2005, beeckmann2017}. Due to the physics-based approach, the asymptotic model extrapolates reasonably outside the temperature and pressure ranges of the calibration data and thus can be readily implemented in combustion engine simulations \cite{ewald2005, hesse2018, hann2020}. One general limitation of the asymptotic model is its applicability to lean equivalence ratios only, although more complex extensions have been proposed to include rich mixtures \cite{hann2020, seshadri1991structure}.

In this section, we utilize numerical data of a well-validated chemical mechanism for methane-air mixtures to develop advanced data-driven laminar flame speed models. This mechanism consists of 79 species and 1055 reactions and we perform one-dimensional unstretched premixed flame simulations using the FlameMaster software package \cite{pitsch1998} to generate a dataset of $S_L$ values that covers a wide range of temperatures $T$ ($300$ to $\SI[]{1200}{\kelvin}$) and pressures $p$ ($1$ to $\SI[]{40}{\bar}$) at lean-to-stoichiometric equivalence ratios $\phi$ ($0.5$ to $1.0$). Extra data points at $\phi = 0.0$ are added to prevent unphysical non-zero flame speeds. The complete dataset of size $m = 9945$ is non-dimensionalized using quantities from the asymptotic analysis of the flame structure, which is essential for developing models that generalize outside the training data distribution.

\subsubsection*{Modeling and training strategies}

The gradient-informed GEP framework is applied to develop analytical models $\hat{S}_L = f\left(\hat{T}, \hat{p}, \hat{\phi} \right)$, where $\hat{\sdot}$ denotes non-dimensionalized quantities. We utilize ten adaptive symbols, which are optimized with the BFGS algorithm\footnote{The LM optimizer without $L_2$ regularization tends to converge to very high numerical constants for regression problems (see Section \ref{sec:rans}) and this section's goal is investigating the impact of the model complexity metric independent from $L_2$ regularization.}, and compare the performance to the standard training approach with random numerical constants (RNC). The available mathematical operators are $+$, $-$, $\times$, $\exp$, $\log$ and $\sqrt{\sdot}$. The evolved models are evaluated using the normalized MSE between the model predictions and the simulation data

\begin{equation}
    J = \frac{1}{m} \sum_{i=1}^m \left(\frac{\hat{S}_{L,i,\text{GEP}} - \hat{S}_{L,i,\text{data}}}{\hat{S}_{L,i,\text{data}}}\right)^2 \, .
\end{equation}

\noindent Additionally, in the second part of the $S_L$ modeling analysis, we investigate setting the model complexity metric $J_c$ (see Eq. \eqref{eq:complexity}) as a second objective function. All others details on the modeling and training settings are listed in Table \ref{tab:strategies}.

\subsubsection*{Analysis and discussion}

\begin{figure}[!ht]
    \centering
    \includegraphics[width=1.0\textwidth]{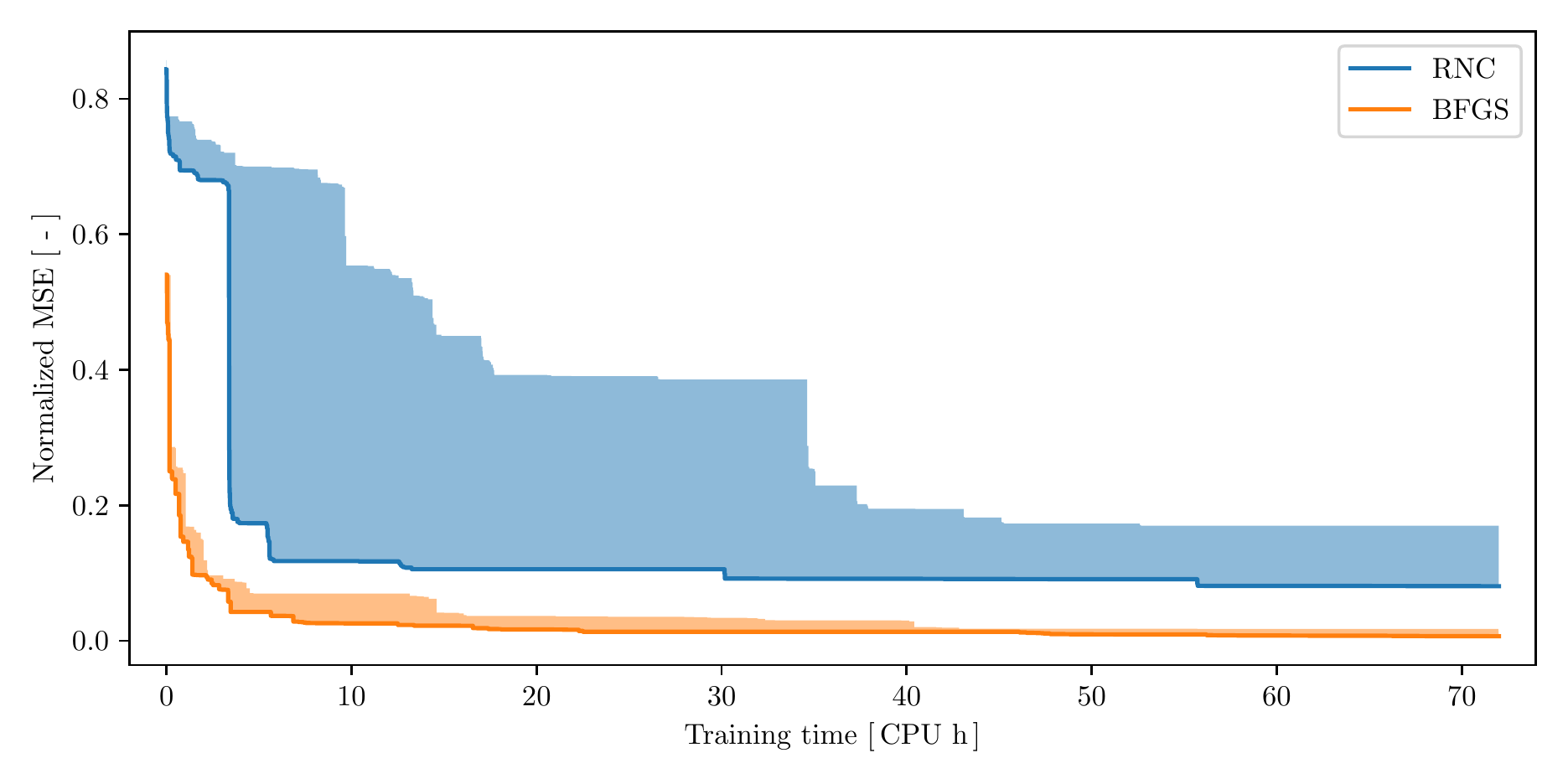}
    \caption{Convergence of normalized MSE of $S_L$ models from standard (RNC) and gradient-informed (BFGS) training without regularization (shaded areas represent variation due to random initialization).}
    \label{fig:lfsconvergence}
\end{figure}

Initially, the $S_L$ models are trained without regularization. Fig. \ref{fig:lfsconvergence} shows the convergence of the normalized MSE objective function for training with adaptive symbols (BFGS) and RNCs. Note that the solid lines represent the minimum error across three randomly initialized training runs and the shaded areas indicate the respective variation. Utilizing adaptive symbols is clearly beneficial compared to the standard training approach, as the normalized MSE is reduced by an entire order of magnitude. Additionally, the variance in the training performance across different random initializations is reduced substantially. The fittest $S_L$ models resulting from the two training approaches are the following:

\begin{align}
    \begin{split}
        \hat{S}_L^\text{RNC} &=  0.91 \cdot \hat{\phi}^2 \cdot (1.10 \cdot \hat{p} + 2.10) \cdot \exp(\hat{p} \cdot \hat{\phi} - \hat{T} + \sqrt{1618.18 \cdot \hat{T}}) \label{eq:lfsrncunreg} \\
        & \cdot (\hat{p} + \hat{\phi} + \sqrt{\hat{T}} - \log(\hat{p}) + 0.43 \cdot \log(|\hat{\phi} + \sqrt{\hat{T}} - 0.09|) + 0.33) \, ,
    \end{split} \\
    \begin{split}
        \hat{S}_L^\text{BFGS} &= \hat{T} \cdot \hat{\phi} \cdot (486.95 \cdot (\hat{p} - 1.30) \cdot (\hat{T} - 0.035) \cdot\sqrt{\hat{p}} - 10.13) \label{eq:lfsbfgsunreg} \\
        & \cdot (-12.49 \cdot \hat{p} + 221923.13 \cdot \hat{T}^2 + 22.94 \cdot (\log(|\hat{\phi} - 0.029|) - \hat{T}) \\
        & + \log(\hat{\phi}) + 153.58) \cdot (\log(|(\hat{p} + \hat{\phi}) \cdot \log(\hat{\phi})|) - 2.07) \, .
    \end{split}
\end{align}

\noindent We observe that both $S_L$ models contain nested terms of the nonlinear operators $\exp$, $\log$ and $\sqrt{\sdot}$, which is considered unphysical and could impair the prediction accuracy when applying the models outside the training data.

\begin{figure}[!ht]
    \centering
    \includegraphics[width=1.0\textwidth]{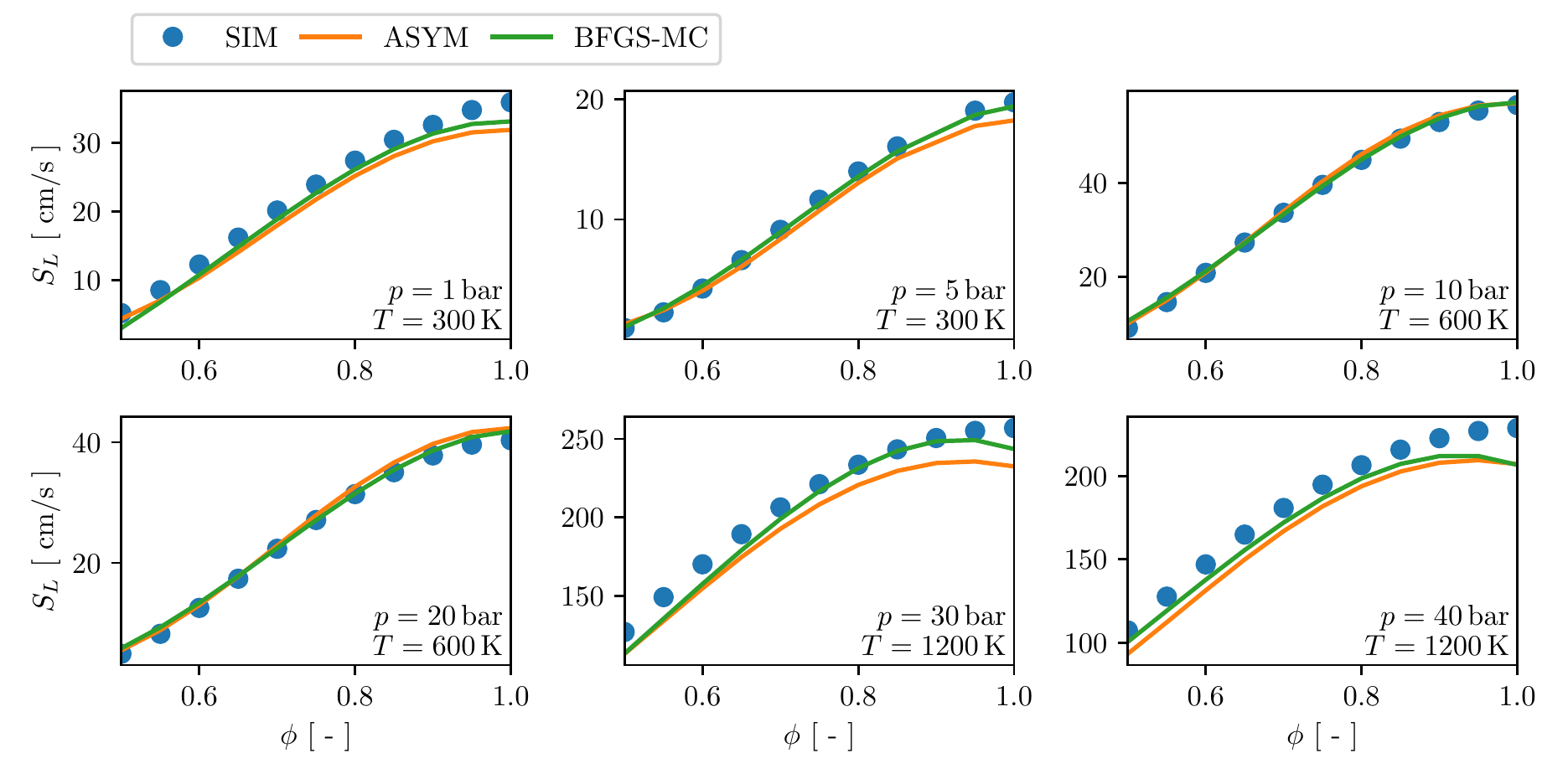}
    \caption{Overview of $S_L$ predictions over equivalence ratio $\phi$ of gradient-informed and regularized model (BFGS-MC) and asymptotic model (ASYM) with target simulation data (SIM) as reference at various pressure and temperature values.}
    \label{fig:lfspreds}
\end{figure}

Thus, we add the model complexity metric $J_c$ as a second objective function and specifically set the symbol complexity values for $s \in \{\exp, \log, \sqrt{\sdot}\}$ from the unity default value to $c_s = 5$, which disincentivizes the usage of these operators. To further reduce the normalized MSE of the developed models, the maximum training runtime is increased from 72 to 120 CPU hours. The resulting $S_L$ model, which is selected as the optimal trade-off between the two objective functions (see Section \ref{sec:rans} for details on the selection process), yields a normalized MSE value of $4.80 \times 10^{-3}$ compared to $6.63 \times 10^{-3}$ for Eq. \eqref{eq:lfsbfgsunreg} and reads as follows:

\begin{equation}
    \begin{split}
    \hat{S}_{L}^\text{BFGS-MC} &= (\hat{T} + 0.008) \cdot (\hat{\phi} - 0.071) \cdot (1.27 \cdot \hat{p} - 27.48 \cdot \hat{T} - 1.18) \cdot 10^6 \label{eq:lfsbfgsreg}  \\
    & \cdot (2.16 \cdot \hat{p} + 55.94 \cdot \hat{T} + 61.26 \cdot \hat{\phi} - \log(\hat{p}) - 3.89) \cdot \sqrt{\hat{T}} \cdot \sqrt{\hat{\phi}} \\
    & \cdot (\hat{T} + \hat{\phi} - 0.034) \, .   
    \end{split}
\end{equation}

\noindent While the overall expression length is similar compared to Eq. \eqref{eq:lfsrncunreg} and \eqref{eq:lfsbfgsunreg}, no nested nonlinear operators occur in Eq. \eqref{eq:lfsbfgsreg}. An overview of the $S_L$ predictions of the developed model (BFGS-MC) at representative $p$ and $T$ values in comparison to the asymptotic model (ASYM) by \citet{gottgens1992} is presented in Fig. \ref{fig:lfspreds}. The prediction accuracy of $\hat{S}_{L}^\text{BFGS-MC}$ improves noticeably over the asymptotic model, in particular towards high equivalence ratios. Therefore, we summarize that the gradient-informed GEP framework distinctly outperforms the standard training approach for this use case and derives a model exceeding the popular asymptotic model. Additionally, model complexity regularization with custom symbol complexities allows steering the model development towards desired model structures.

\subsection{Subgrid-scale modeling}
\label{sec:les}

Large eddy simulations are one of the most important methods for turbulence simulation and have been widely applied in various areas, e.g. in aerospace engineering. LES obtain large-scale flow structures by solving filtered Navier-Stokes equations, while the effects of subgrid-scale structures on the large-scale structures are approximated by SGS models \cite{smagorinsky1963,lilly1967,pope2001,sagaut2006}. As current models do not universally produce accurate predictions, improving the modeling of the SGS terms remains a focus of LES studies.

For this use case, we consider incompressible turbulence for which the filtered continuity and momentum equations \cite{sagaut2006,meneveau2000} are 

\begin{align}
    \frac{\partial{\widetilde{u}}_{i}}{\partial x_{i}} &= 0 \, , \label{eq:continuity} \\
    \frac{\partial{\widetilde{u}}_{i}}{\partial t} + \frac{\partial{\widetilde{u}}_{i}{\widetilde{u}}_{j}}{\partial x_{j}} &= - \frac{\partial\widetilde{p}}{\partial x_{i}} - \frac{\partial\tau_{ij}}{\partial x_{j}} + \nu\frac{\partial^{2}{\widetilde{u}}_{i}}{\partial x_{j}x_{j}} + {\widetilde{\mathcal{F}}}_{i} \, , \label{eq:momentum}
\end{align}

\noindent where $\widetilde{\cdot}$ denotes filtered variables and is defined for the velocity $u$ as 

\begin{equation}
    \widetilde{u}(x,t) = \int_{D} G(r,x) u(x-r,t) dr \, ,
\end{equation}

\noindent where $D$ is the entire flow domain and $G$ is the filter function which satisfies the normalization condition $\int G(r,x) dr = 1$. The SGS stress tensor $\tau_{ij}$ in Eq. \eqref{eq:momentum} is defined as $\tau_{ij} = \widetilde{u_{i}u_{j}} - {\widetilde{u}}_{i}{\widetilde{u}}_{j}$.

In this section, we apply the gradient-informed GEP framework to develop improved SGS models. The high-fidelity dataset for training and testing is obtained from a direct numerical simulation (DNS) of three-dimensional forced incompressible isotropic turbulence \cite{xie2020}. The DNS velocity field is calculated on a uniform grid of size $1024^{3}$ and the Taylor Reynolds number is $Re_{\lambda} \approx 260$. To extract the filtered velocity field, we apply a top-hat filter in three dimensions. The top-hat filter in one dimension is

\begin{equation}
    {\widetilde{f}}_{i} = \frac{1}{2n_f}\left( {f_{i - n_f/2} + 2{\sum\limits_{j = i - \frac{n_f}{2} + 1}^{i + \frac{n_f}{2} - 1}{f_{j} + f_{i + n_f/2}}}}\right) \, ,
\end{equation}

\noindent where the filter width can be expressed as $\Delta = n_f \Delta x$. We choose $n_f = 16$ in this case such that about $5\%$ of the turbulence kinetic energy is filtered, as discussed by \citet{xie2020}. Furthermore, coarse-graining that yields $16^{3}$ different datasets with $64^{3}$ coarse grid points each is applied to the filtered DNS data. We select a total dataset of $8 \times 64^{3}$ grid points and $70\%$ of points are randomly selected for training and the remaining $30\%$ for testing.

\subsubsection*{Modeling and training strategies}

For this use case, we compare the BFGS and LM optimizers in combination with $L_2$ regularization for modeling the anisotropic SGS stress tensor $\tau_{ij}^{A}$. Following \citet{li2021}, we model $\tau_{ij}^{A}$ as a function of the filter width $\Delta$ and the local filtered strain and rotation rate tensors ${\widetilde{S}}_{ij}$ and ${\widetilde{\Omega}}_{ij}$. We assume a linear dependency on $\Delta$ and utilize the integrity basis proposed by \citet{pope1975} to model $\tau_{ij}^{A}$ as 

\begin{equation}
    \tau_{ij}^{A} = (\Delta |\widetilde{S}|)^{2} \cdot \sum\limits_{k = 1}^4 g^k\left( {I^1, \ldots ,I^4} \right) V_{ij}^k \, ,
\end{equation}

\noindent where $|\widetilde{S}| = \sqrt{{\widetilde{S}}_{mn}{\widetilde{S}}_{mn}}$ is an inverse time scale. The basis tensors $V_{ij}^k$ and the invariants $I^l$ are only functions of ${\widetilde{S}}_{ij}$ and ${\widetilde{\Omega}}_{ij}$ non-dimensionalized by $|\widetilde{S}|$. The definitions of $V_{ij}^k$ and $I^l$ are available in \cite{li2021}. The scalar functions $g^k$ are the modeling target of the GEP framework.

During training, the MSE between the predicted anisotropic SGS stress tensor $\tau_{ij}^{A,\text{GEP}}$ and the DNS training data value $\tau_{ij}^{A,\text{data}}$ is minimized. Therefore, a population of 100 individuals with four adaptive symbols per scalar function is evolved. Table \ref{tab:strategies} lists the remaining modeling and training settings.

\subsubsection*{Analysis and discussion}

\begin{figure}[!ht]
    \centering
    \includegraphics[width=1.0\textwidth]{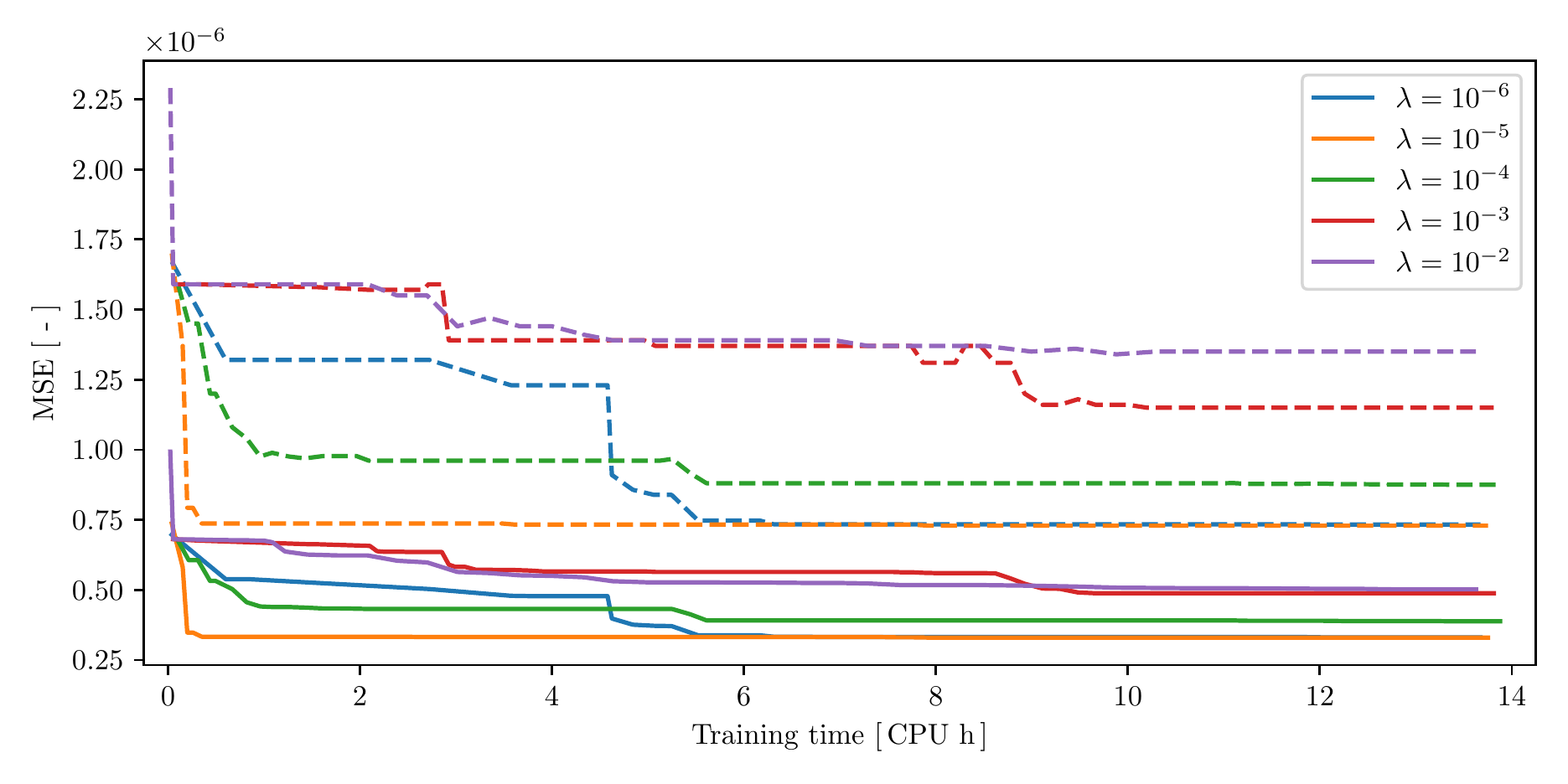}
    \caption{Convergence of MSE of SGS models in $\lambda$ hyperparameter study evaluated on training (solid) and testing (dashed) data.}
    \label{fig:l2hyperparameter}
\end{figure}

First, we perform a hyperparameter study to identify an appropriate order of magnitude for the $L_2$ regularization parameter $\lambda$ in Eq. \eqref{eq:l2regularization}. We utilize the BFGS optimizer and explore $\lambda \in \{10^{-2}, 10^{-3}, 10^{-4}, 10^{-5}, 10^{-6}\}$. The maximum training runtime is set to 13.89 CPU hours. 

Fig. \ref{fig:l2hyperparameter} shows the MSE of SGS models regularized with different $\lambda$ values during training. There is approximately a constant factor between the training and testing MSE across all $\lambda$ values, displayed as solid and dashed lines, respectively. Improvements on the training dataset result in improvements on the testing dataset, which suggests that no overfitting occurs. Furthermore, we observe that starting with $\lambda = 10^{-5}$ the MSE increases with increasing $\lambda$ value. $L_2$ regularization incentivizes parameters of small magnitude and thus, restricts the search space for adaptive symbol values. From $\lambda = 10^{-4}$, the search space gets too restricted and the optimal numerical constants cannot be learned, which results in deteriorating model performances. However, comparing the results based on $\lambda = 10^{-5}$ and $\lambda = 10^{-6}$ indicates that a limited amount of $L_2$ regularization, i.e. a small restriction to the parameter search space, improves the convergence speed significantly. 

\begin{figure}[!ht]
    \centering
    \includegraphics[width=1.0\textwidth]{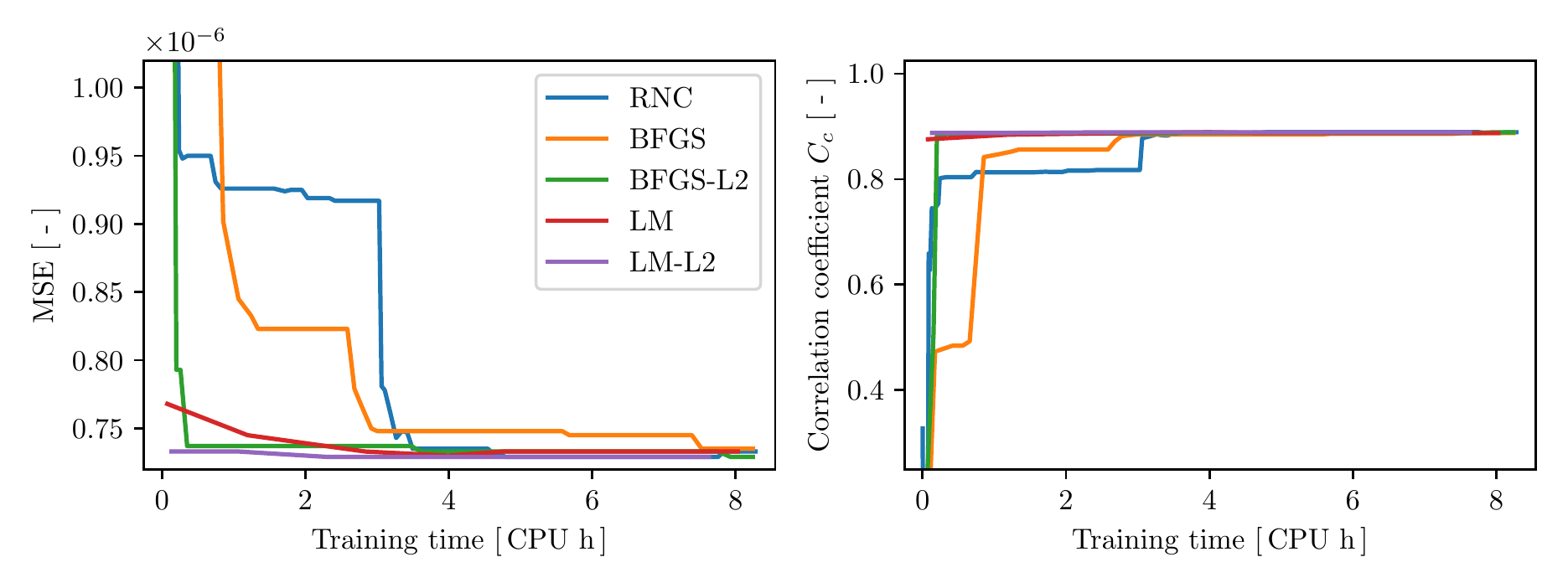}
    \caption{Convergence of MSE (left) and $C_c$ (right) of different SGS models evaluated on testing data.}
    \label{fig:sgsconvergence}
\end{figure}

In the following, we select $\lambda = 10^{-5}$ and compare the BFGS and LM optimizers, both with and without $L_2$ regularization, to the standard GEP training approach. We analyze the convergence of the MSE and the correlation coefficient $C_c$ in Fig. \ref{fig:sgsconvergence} for the different approaches evaluated on the testing dataset. The correlation coefficient measures the average componentwise correlation between the predicted and the high-fidelity anisotropic SGS stress tensors $\tau_{ij}^{A,\text{GEP}}$ and $\tau_{ij}^{A,\text{data}}$ and is calculated as 

\begin{equation}
    C_c = \frac{1}{6} \sum_{i=1}^3 \sum_{j=i}^3 \frac{\left\langle \left(\tau^{\text{GEP}} - \left\langle \tau^{\text{GEP}} \right\rangle \right) \odot \left(\tau^{\text{data}} - \left\langle \tau^{\text{data}} \right\rangle \right) \right\rangle}{\sqrt{\left\langle \left(\tau^{\text{GEP}} - \left\langle \tau^{\text{GEP}} \right\rangle \right)^2 \right\rangle  \odot \left\langle \left(\tau^{\text{data}} - \left\langle \tau^{\text{data}} \right\rangle \right)^2 \right\rangle}} \, ,
\end{equation}

\noindent where $\left\langle \cdot \right\rangle$ indicates averaging over the dataset, $\odot$ represents componentwise multiplication and $\tau$ is a short notation for $\tau_{ij}^A$.

Fig. \ref{fig:sgsconvergence} (left) shows that all training approaches converge to a similar MSE value within the training runtime of 8.3 CPU hours. This can be explained by the simple model structure of the fittest SGS models. Using the LM optimizer with $L_2$ regularization, the resulting model for $\tau_{ij}^{A,\text{GEP}}$ is

\begin{equation}
    \tau_{ij}^{A,\text{GEP}} = (\Delta |\widetilde{S}|)^{2} \cdot \left(- 0.01V_{ij}^1 - 0.1V_{ij}^2 + 0.07V_{ij}^3 - 0.11V_{ij}^4\right) \, ,
    \label{eq:sgsmodel}
\end{equation}

\noindent where the scalar functions $g^k$ are constant. Thus, determining accurate numerical constants is the key challenge for this use case. Consequently, utilizing adaptive symbols and numerical optimizers significantly increases the convergence speed, although the BFGS optimizer without $L_2$ regularization appears to converge to suboptimal constants until close to the maximum training time. The general observations are that the least squares-specific LM optimizer outperforms the BFGS optimizer and that $L_2$ regularization further speeds up training convergence, which is in line with the results of Section \ref{sec:sutherland} and the $\lambda$ hyperparameter study. Furthermore, $C_c$ values of close to $0.9$ in Fig. \ref{fig:sgsconvergence} (right) support that the resulting models are highly correlated with the high-fidelity data and that, for example, not only the error in one component of $\tau_{ij}^{A,\text{GEP}}$ is reduced, which might cause unphysical predictions.

\begin{figure}[!ht]
    \centering
    \includegraphics[width=1.0\textwidth]{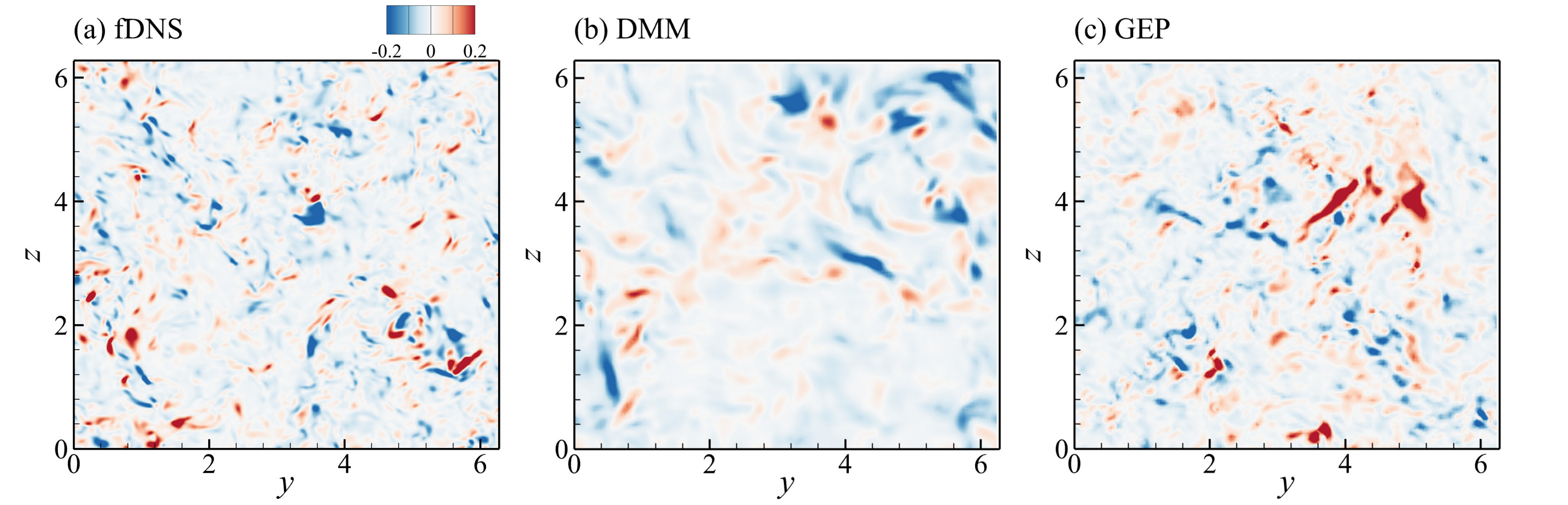}
    \caption{Contours of SGS stress component $\tau_{23}^{A}$ at arbitrary domain slice: (a) fDNS, (b) DMM, (c) GEP.}
    \label{fig:sgsaposteriori}
\end{figure}

In order to investigate the suitability of Eq. \eqref{eq:sgsmodel} as a closure model, i.e. ensuring accurate predictions and numerical stability, we perform an LES of homogeneous isotropic turbulence with the developed SGS model. The LES grid is of size $128^3$ with a filter width of $\Delta = 16 \Delta x$ and the numerical methods are described in \cite{li2021}. We compare the LES results based on $\tau_{ij}^{A,\text{GEP}}$ with the filtered high-fidelity DNS data (fDNS) and the predictions of an LES applying a standard SGS model, the dynamic mixed model (DMM) \cite{liu1994}. 

Fig. \ref{fig:sgsaposteriori} displays the $\tau_{23}^{A}$ contours of the different simulations at an arbitrary slice in the flow domain. The SGS model developed with the gradient-informed GEP framework captures fine-scale structures similar to the fDNS data and predicts stress intensity levels accurately. In contrast, the DMM models more dissipation which leads to larger $\tau_{23}^{A}$ structures of lower intensity.

In summary, adaptive symbols, especially when employing the LM optimizer, and a moderate level of $L_2$ regularization significantly increase the convergence speed when developing models with a simple structure such as in Eq. \eqref{eq:sgsmodel}. The developed SGS model is easily implementable in an LES solver and was demonstrated to substantially improve $\tau_{ij}^{A}$ predictions over a standard SGS model.

\subsection{Nonlinear eddy viscosity modeling}
\label{sec:rans}

RANS calculations remain the primary tool to perform CFD simulations of turbulent flows of industrial interest. In contrast to spatial filtering in LES, Reynolds averaging allows calculating time-averaged flow fields, which is computationally more efficient and often sufficient for industrial applications. For an incompressible flow with constant density, the Reynolds-averaged continuity and momentum equations are

\begin{align}
    \frac{\partial\overline{u}_i}{\partial x_i} &= 0 \, , \\
    \overline{u}_j \frac{\partial\overline{u}_i}{\partial x_j} &= - \frac{\partial\overline{p}}{\partial x_i} + \frac{\partial}{\partial x_j} \left( \nu \frac{\partial\overline{u}_i}{\partial x_j} - \overline{u^\prime_i u^\prime_j} \right) \, ,
\end{align}

\noindent with the mean velocity $\overline{u}_i$, the density-corrected pressure $\overline{p}$ and the kinematic viscosity $\nu$. The Reynolds stress tensor $\overline{u^\prime_i u^\prime_j}$ describes the impact of turbulent fluctuations on the averaged flow field and can be deconstructed into an isotropic ($\frac{2}{3} k \delta_{ij}$) and an anisotropic term ($2 k a_{ij}$), where $k$ is the turbulence kinetic energy and $\delta_{ij}$ is the Kronecker delta. 

The anisotropy tensor $a_{ij}$ is modeled as $a_{ij} = -\frac{\nu_t}{k} S_{ij}$ by linear eddy viscosity models (LEVM), such as the popular $k$-$\omega$ SST model \cite{menter1994}, which solves two additional transport equations for $k$ and $\omega$, the specific dissipation rate, to calculate the eddy viscosity $\nu_t$. The mean strain rate $S_{ij}$ is calculated from $S_{ij} = \frac{1}{2} \left(\frac{\partial\overline{u}_i}{\partial x_j} + \frac{\partial\overline{u}_j}{\partial x_i}\right)$. To improve the well-known shortcomings of LEVMs, e.g. predicting flows with separation or curvature inaccurately \cite{leschziner2015}, NLEVMs assume a nonlinear dependency of $a_{ij}$ on $S_{ij}$ and the mean rotation rate $\Omega_{ij} = \frac{1}{2} \left(\frac{\partial\overline{u}_i}{\partial x_j} - \frac{\partial\overline{u}_j}{\partial x_i}\right)$.

In this section, we develop NLEVMs for a wall-mounted square cylinder flow using the GEP framework with adaptive symbols and the introduced regularization techniques. In addition to modeling $a_{ij}$, we evolve a turbulence production correction model $R$, as proposed by \citet{schmelzer2019}. This second model extends the production term of $k$ in the turbulence transport equations and ensures the consistency of $k$ between its transport equation and its high-fidelity data values.

This section's use case is a complex three-dimensional flow around a square cylinder with a height-to-width ratio of $h/d = 4$ at a Reynolds number of $Re_d = 11,000$. The resulting flow features include horseshoe vortices around the cylinder, upwash from the boundary layer, downwash over the cylinder tip and a von Kármán vortex street \cite{wang2009}. As high-fidelity data for developing the $a_{ij}$ and $R$ models, we utilize a well-validated hybrid RANS/LES dataset generated by \citet{weatheritt2016a}. Following the results of \citet{haghiri2020}, who split the flow domain of this case into a near-body region and a downstream region and showed that the NLEVM developed for the downstream region is responsible for nearly all prediction improvements, we extract the training and testing data from the downstream region starting at $x/d = 2$. As a compromise between low computational cost and avoiding overfitting, $10^5$ data points are selected for training while the remaining data points are set aside for testing. In order to calculate $\omega$ and $R$ values from the high-fidelity data, we employ the k-corrective frozen RANS approach \cite{schmelzer2019}, which solves the turbulence transport equations with frozen $\overline{u}_i$, $k$ and $a_{ij}$ high-fidelity data values.

\subsubsection*{Modeling and training strategies}

For developing $a_{ij}$ and $R$ models, we utilize adaptive symbols and both $L_2$ and model complexity regularization. We analyze the performance of the BFGS and LM optimizers and make a comparison to unregularized training and the standard GEP training approach. The modeling approach for the turbulence production correction $R$ is 

\begin{equation}
    R = 2 k a_{ij}^R \frac{\partial\overline{u}_i}{\partial x_j} \, ,
    \label{eq:productioncorrection}
\end{equation}

\noindent where $a_{ij}^R$ is modeled similar to $a_{ij}$ based on the integrity basis derived by \citet{pope1975}, which is defined for the anisotropy tensor as 

\begin{equation}
    a_{ij} = \sum\limits_{k=1}^{10} g^k\left(I^1, I^2, ..., I^5\right)V_{ij}^k \, .
\label{eq:integritybasis}
\end{equation}

\noindent In contrast to Section \ref{sec:les}, all five scalar invariants $I^l$ and ten basis tensors $V_{ij}^k$ are included to capture the complexity of this use case. The mean strain and rotation rates $S_{ij}$ and $\Omega_{ij}$ are non-dimensionalized by $\omega$ before calculating $I^l$ and $V_{ij}^k$ according to the definitions in \cite{pope1975}. The scalar functions $g^k$ are learned by the GEP framework.

The training objective is minimizing the MSE between the predictions of the $a_{ij}$ and $R$ models and the corresponding training data values. For $L_2$ regularization, the MSE objective function is extended according to Eq. \eqref{eq:l2regularization}. The regularization parameter is determined via a hyperparameter study to $\lambda = 10^{-7}$ (see \ref{sec:hyperparameters}). For model complexity regularization, Eq. \eqref{eq:complexity} is set as a second objective function. To further support the development of low complexity models, the numerical constant $0$ is assigned a symbol complexity of $c_0 = 0$, as, for example, multiplication with $0$ reduces complexity beyond the calculated $J_c$ value. All other symbols $\mathcal{S} \setminus \{0\}$ have a symbol complexity of unity.

During training, a population of 100 individuals is evolved for a maximum training runtime of 240 CPU hours. The optimal number of adaptive symbols is a trade-off between functional expressivity and computational cost, as one adaptive symbol is insufficient to approximate a function with multiple distinct numerical constants, but more adaptive symbols complicate the optimization problem. In \ref{sec:hyperparameters}, we identify five adaptive symbols per scalar function $g^k$ to be an optimal trade-off. The remaining modeling and training settings are listed in Table \ref{tab:strategies}.

\subsubsection*{Analysis and discussion}

\begin{figure}[!ht]
    \centering
    \includegraphics[width=1.0\textwidth]{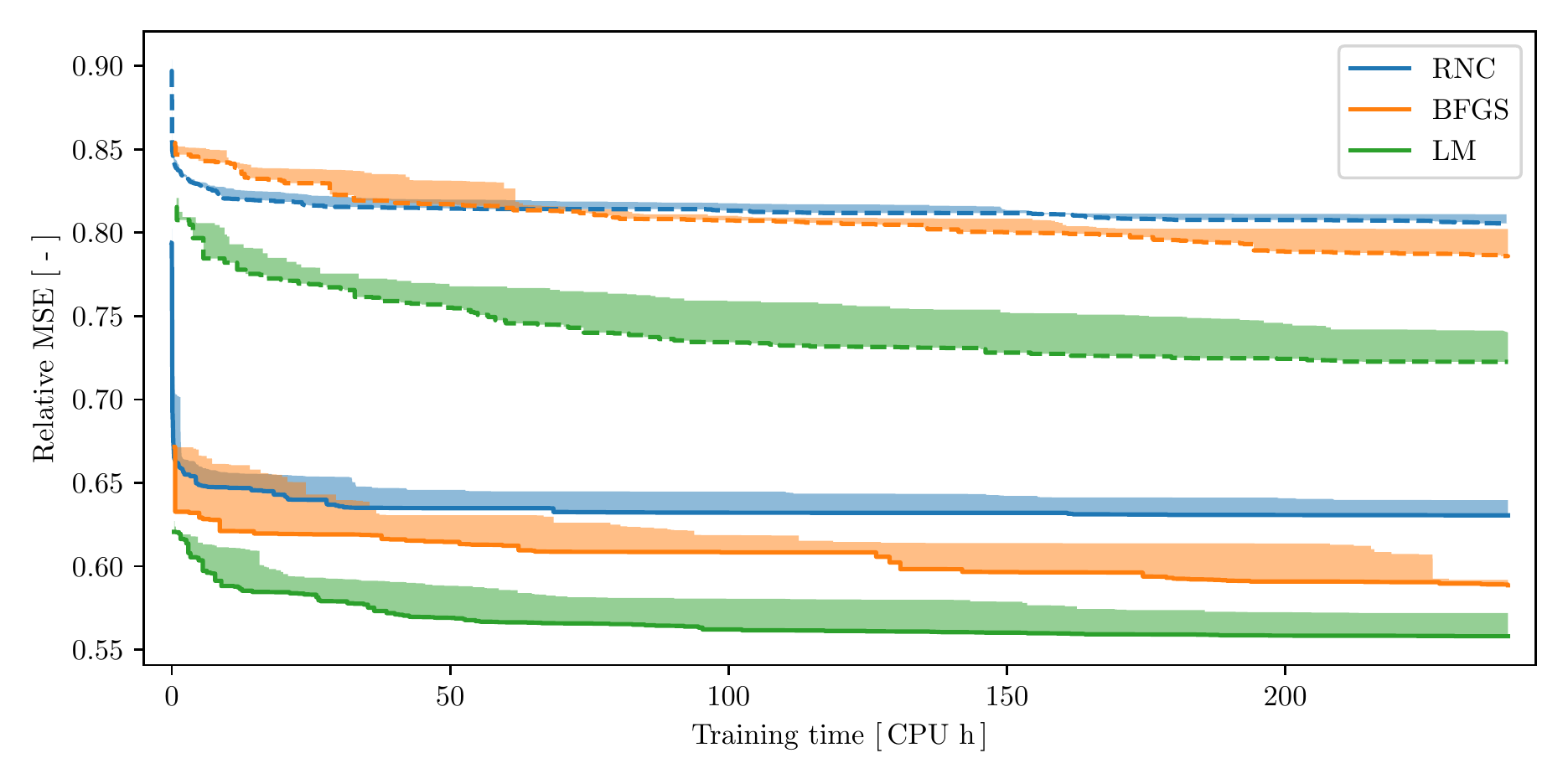}
    \caption{Convergence of MSE relative to LEVM of $a_{ij}$ (dashed) and $R$ (solid) models trained without regularization for NLEVM case (shaded areas represent variation due to random initialization).}
    \label{fig:nlevmconvergence}
\end{figure}

First, we develop $a_{ij}$ and $R$ models without regularization. Fig. \ref{fig:nlevmconvergence} shows the convergence of the MSE, relative to the linear $k$-$\omega$ SST model, of the different training approaches, where shaded areas represent variation resulting from different random initializations. All training approaches improve both models over the LEVM and the relative improvement is larger for the turbulence production correction model $R$. Utilizing adaptive symbols is clearly advantageous in comparison to the standard training approach (RNC) and, in line with the results of Sections \ref{sec:sutherland} and \ref{sec:les}, the LM optimizer yields lower MSE values than the BFGS optimizer.

\begin{table}[ht]
	\centering
	\caption{Training and testing MSE relative to LEVM, model complexity $J_c$ and maximum absolute adaptive symbol value $\max_i(|p_i|)$ of $a_{ij}$ and $R$ models trained without regularization for NLEVM case.}
    \begin{tabular}{l l c c c c}
        \toprule
        Model & Strategy & Train. MSE & Test. MSE & $J_c$ & $\max_i(|p_i|)$ \\
        \cmidrule(lr){1-6}
        \multirow{3}{*}{$a_{ij}$}   & RNC   & 0.8056 & 0.8049 & 330 & --  \\
                                    & BFGS  & 0.7858 & 0.7849 & 140 & $1.09 \times 10^2$ \\
                                    & LM    & 0.7226 & 0.7224 & 180 & $3.71 \times 10^{14}$ \\
        \cmidrule(lr){1-6}                                    
        \multirow{3}{*}{$R$}        & RNC   & 0.6305 & 0.6189 & 320 & --  \\
                                    & BFGS  & 0.5886 & 0.5789 & 166 & $4.24 \times 10^2$ \\
                                    & LM    & 0.5581 & 0.5497 & 210 & $3.06 \times 10^{13}$ \\
        \bottomrule
    \end{tabular}
	\label{tab:nlevmunregularized}
\end{table}

Table \ref{tab:nlevmunregularized} presents the performance of the resulting $a_{ij}$ and $R$ models on the training and testing datasets. Additionally, the respective model complexity $J_c$ and the maximum adaptive symbol magnitude $\max_i(|p_i|)$ are listed, which are both not regularized in this first study. Comparing the errors on the training and testing datasets shows that no overfitting occurs. However, the model complexity values of the developed models are too high to allow interpretation. For example, the following is the $a_{ij}$ model developed using the BFGS optimizer, which is the least complex trained model and has additionally been simplified using the SymPy package \cite{meurer2017}:

\begin{align*}
    a_{ij}^\text{BFGS} = \;&(306.79 \cdot I^3 + (0.88 - 175.64 \cdot I^2)\cdot(-I^1 + I^3 + 0.12) - 1.86) \cdot V_{ij}^1 \\
    & + (I^2 - 0.02)\cdot(I^5 + 109.19) \cdot V_{ij}^2 - (-I^1 + I^2 + I^3 + 2.54) \cdot V_{ij}^3 \\ 
    & + (I^1 - 15.42)\cdot(I^1 + I^2 + 0.08)\cdot(84.03 \cdot I^2 - 84.03 \cdot I^4 + 6.74) \cdot V_{ij}^4 \\
    & + (I^3 \cdot (I^1 + I^3) + 12.91) \cdot V_{ij}^5 + I^3 \cdot (I^2 + 263.97)\cdot(I^4 + 41.48) \cdot V_{ij}^6 \\
    & - 191.44\cdot (I^1 - 1) \cdot V_{ij}^7 + (5.15 \cdot I^1 + I^2 + 36.44) \cdot V_{ij}^8 \\
    & + (I^3 + I^5 - 3.07) \cdot V_{ij}^9 + (I^4 + 2.13) \cdot V_{ij}^{10} \, .
\end{align*}

The standard training approach (RNC) yields the models with the highest complexity values $J_c$. In order to approximate a certain numerical constant, the combination of multiple random numerical constants and mathematical operators is generally required. This increases the search space of possible expressions and complicates symbolic regression. The gradient-informed GEP framework derives an accurate numerical constant with a single adaptive symbol. However, we notice that the LM optimizer converges to extraordinarily high $p_i$ values. Table \ref{tab:nlevmunregularized} shows that the LM-optimized models do not overfit the training data, but numerical constants on the order of $\mathcal{O}(10^{13})$ will likely cause stability issues of the numerical solver that applies the trained models.

\begin{figure}[!ht]
    \centering
    \includegraphics[width=0.75\textwidth]{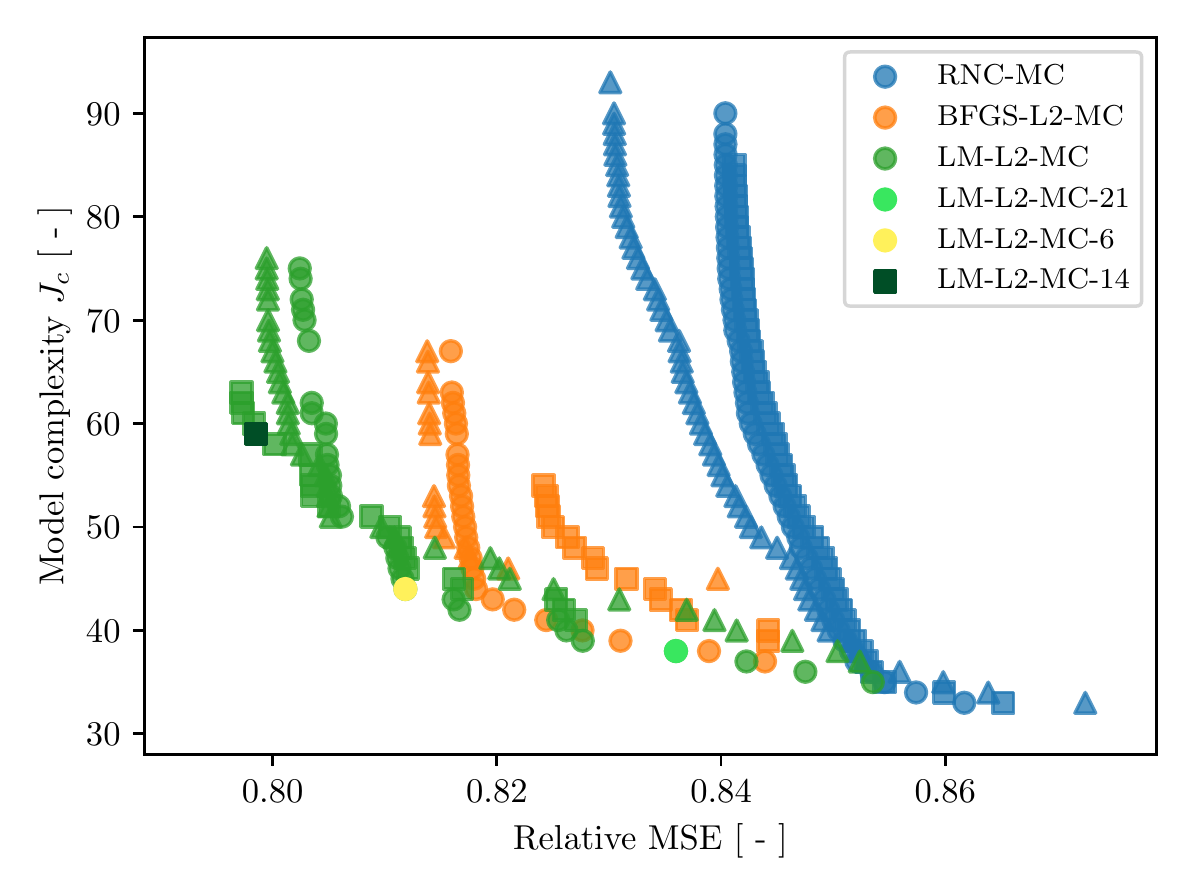}
    \caption{MSE relative to LEVM and model complexity $J_c$ of $a_{ij}$ models developed with regularized training approaches for NLEVM case (markers represent different random initialization).}
    \label{fig:nlevmparetoA}
\end{figure}

\begin{figure}[!ht]
    \centering
    \includegraphics[width=0.75\textwidth]{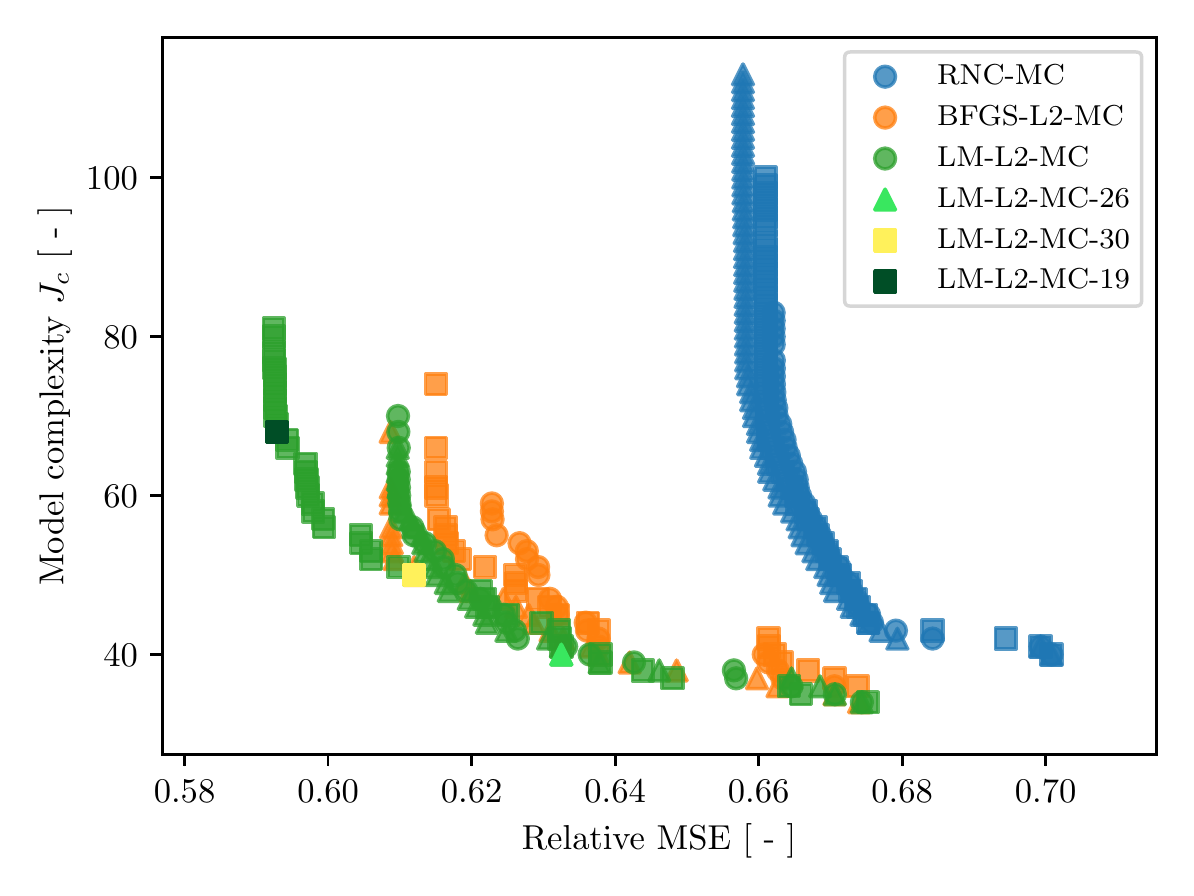}
    \caption{MSE relative to LEVM and model complexity $J_c$ of $R$ models developed with regularized training approaches for NLEVM case (markers represent different random initialization).}
    \label{fig:nlevmparetoR}
\end{figure}

To develop interpretable and implementable closure models, we apply $L_2$ regularization when optimizing adaptive symbol values and add the model complexity metric defined in Eq. \eqref{eq:complexity} as a second objective function. Fig. \ref{fig:nlevmparetoA} and \ref{fig:nlevmparetoR} show the performance of the developed $a_{ij}$ and $R$ models on the two objective functions. Note that the marker symbols represent training runs with different random initializations and that only the Pareto front is plotted, i.e. models that are not outperformed by any other model in the training run population. The advantages of utilizing adaptive symbols compared to the standard training approach (RNC-MC) are significant, as lower prediction errors and $J_c$ values are achieved for both models. We notice that the LM optimizer improves over the BFGS optimizer especially towards higher $J_c$ values, where more adaptive symbols in the expression tree are likely and thus, the optimization problem is more complex. In comparison to unregularized training, the complexity values of the gradient-informed models reduce from a minimum of 140 to a maximum of 81, while the relative MSE values are slightly increased (by 0.076 for $a_{ij}$ and 0.035 for $R$).

One benefit of multi-objective optimization is that the trade-off between the objective functions can be done after the training. We select three candidate models of increasing complexity for $a_{ij}$ and $R$, which are highlighted in light green, yellow and dark green in Fig. \ref{fig:nlevmparetoA} and \ref{fig:nlevmparetoR}. Table \ref{tab:nlevmregularized} demonstrates that, in addition to avoiding overfitting and reducing model complexity, the adaptive symbol values are on the order of $\mathcal{O}(10^1)$ across the varying $J_c$ values. Therefore, we conclude that the two regularization techniques are effective.

\begin{table}[ht]
	\centering
	\caption{Training and testing MSE relative to LEVM, model complexity $J_c$ and maximum absolute adaptive symbol value $\max_i(|p_i|)$ of $a_{ij}$ and $R$ models trained with regularization for NLEVM case.}
    \begin{tabular}{l l c c c c}
        \toprule
        Model & Strategy & Train. MSE & Test. MSE & $J_c$ & $\max_i(|p_i|)$ \\
        \cmidrule(lr){1-6}
        \multirow{3}{*}{$a_{ij}$}   & LM-L2-MC-21 & 0.8360 & 0.8352 & 38 & $1.15 \times 10^1$ \\
                                    & LM-L2-MC-6  & 0.8118 & 0.8109 & 44 & $1.27 \times 10^1$ \\
                                    & LM-L2-MC-14 & 0.7985 & 0.7978 & 59 & $1.30 \times 10^1$ \\
        \cmidrule(lr){1-6}                                    
        \multirow{3}{*}{$R$}        & LM-L2-MC-26 & 0.6325 & 0.6212 & 40 & $1.37 \times 10^1$ \\
                                    & LM-L2-MC-30 & 0.6120 & 0.6012 & 50 & $1.86 \times 10^1$ \\
                                    & LM-L2-MC-19 & 0.5929 & 0.5828 & 68 & $1.64 \times 10^1$ \\
        \bottomrule
    \end{tabular}
	\label{tab:nlevmregularized}
\end{table}

For a detailed analysis, the models LM-L2-MC-6 ($a_{ij}^{6}$) and LM-L2-MC-30 ($R^{30}$) are selected as a compromise between prediction accuracy and model complexity. The respective model expressions are the following:

\begin{align}
    \begin{split}
    a_{ij}^{6}  = \;&(160.48 \cdot I^3 - 1.41) \cdot V_{ij}^1 + (89.48 \cdot I^2 - 1.41) \cdot V_{ij}^2  \label{eq:nlevmA} \\
            &- 6.95 \cdot V_{ij}^4 + 159.33 \cdot V_{ij}^7 + 111.25 \cdot V_{ij}^9 \, , 
    \end{split} \\
    \begin{split}
    R^{30}       = \;&2 k \cdot \left( \left(-101.10 \cdot I^1 + 236.20 \cdot I^2 + 7.19\right) \cdot V_{ij}^1 \right. \label{eq:nlevmR} \\
            &- \left.\left(5.86 \times 10^4 \cdot I^1 I^2 - 1.09 \times 10^4 \cdot I^2\right) \cdot V_{ij}^6 \right)  \frac{\partial\overline{u}_i}{\partial x_j} \, ,
    \end{split}
\end{align}

\noindent which reduce the number of basis tensors $V_{ij}^k$ from ten in unregularized training to merely five and two, respectively, and contain mostly constant and linear terms. Interestingly, despite a $\max_i(|p_i|)$ value of 18.6 and applying model complexity regularization, the LM-L2-MC-30 model achieves constants on the order of $\mathcal{O}(10^4)$ by combining multiple adaptive symbols, which signals the importance of the multiplied tensor $V_{ij}^6$.

\begin{figure}[!ht]
    \centering
    \includegraphics[width=0.75\textwidth]{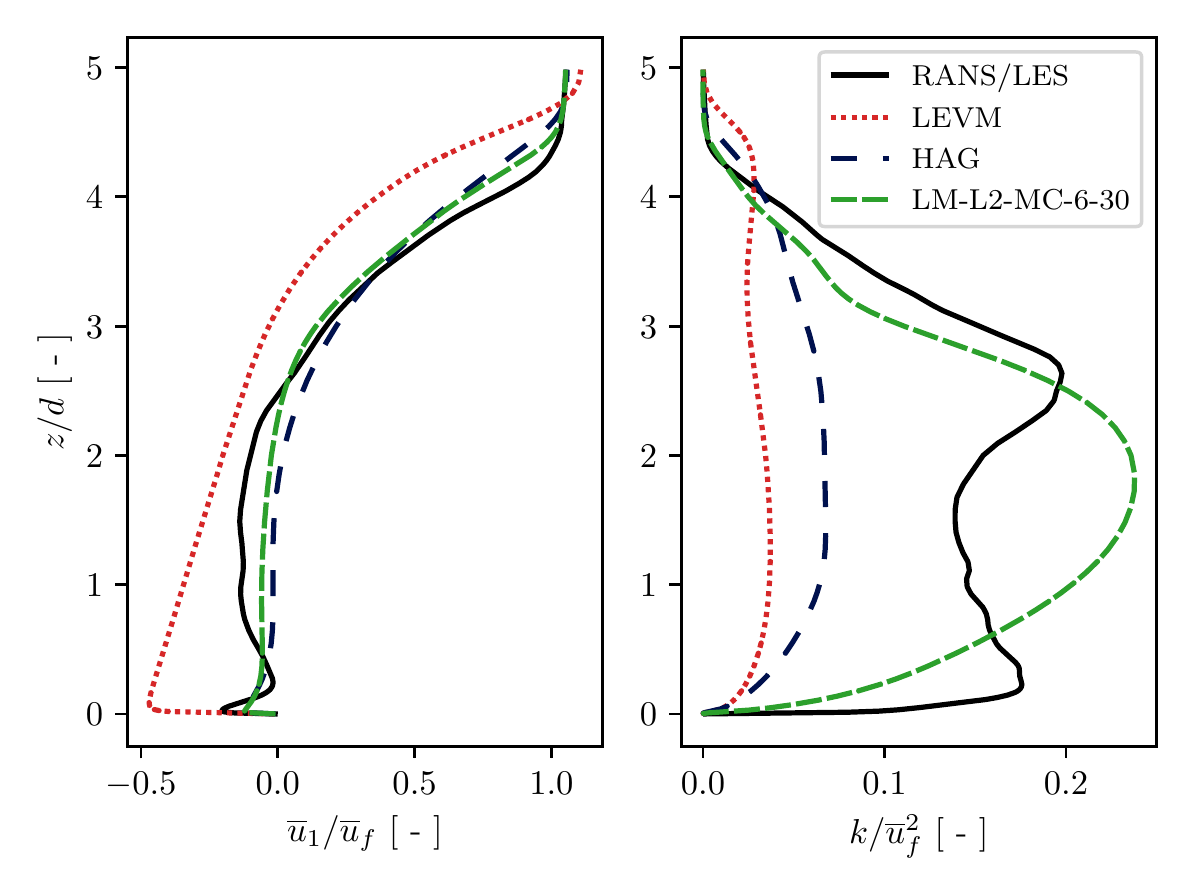}
    \caption{Wall-normal profiles of mean streamwise velocity (left) and turbulence kinetic energy (right) on centerline at $x/d = 3$ of NLEVM case.}
    \label{fig:nlevmprofiles}
\end{figure}

Lastly, we investigate the suitability of the models in Eq. \eqref{eq:nlevmA} and \eqref{eq:nlevmR} as closure models. The models are implemented in the CFD software OpenFOAM \cite{weller1998} to extend the baseline $k$-$\omega$ SST model. We run a steady-state RANS calculation on the numerical grid of the hybrid RANS/LES simulation that generated the high-fidelity data \cite{weatheritt2016a}. Fig. \ref{fig:nlevmprofiles} shows the wall-normal profiles of the mean streamwise velocity $\overline{u}_1$ and the turbulence kinetic energy $k$, non-dimensionalized by the freestream velocity $\overline{u}_f$, on the centerline at $x/d = 3$. We compare the two developed models (LM-L2-MC-6-30) to the high-fidelity hybrid RANS/LES data, the baseline LEVM and the NLEVM developed by \citet{haghiri2020} for the downstream region (HAG). In addition to clear improvements over the LEVM, the LM-L2-MC-6-30 models predict the $\overline{u}_1$ profile slightly more accurately than the HAG model. However, the most noticeable advancement of the LM-L2-MC-6-30 models is predicting the turbulence kinetic energy on level with the high-fidelity data. While some inaccuracies in the $k$ profile remain due to the unsteady nature of the flow features in this use case, the improvements compared to the HAG model are substantial.

We summarize that adaptive symbols lead to significant improvements in unregularized and regularized training with the GEP framework for this complex three-dimensional flow. The two introduced regularization techniques allow the development of implementable and more interpretable $a_{ij}$ and $R$ models, which, for the first time, yield steady-state RANS predictions for $k$ on level with the high-fidelity data.

\section{Conclusion}
\label{sec:conclusion}

The concept of adaptive symbols is introduced in this paper to advance the development of physics closure models from high-fidelity data. The fundamentally stochastic GEP framework by \citet{weatheritt2016} is extended via adaptive symbols to incorporate gradient information in order to learn accurate numerical constants. A general, i.e. objective function independent, optimizer (BFGS) and a specific nonlinear least squares optimizer (LM) are compared to determine locally optimal adaptive symbol values. Furthermore, two regularization methods are implemented to support the development of interpretable and implementable closure models, which is typically associated with numerical constants of small magnitude and model expressions of low complexity. We add $L_2$ regularization to the objective function for the gradient-based optimization and define a novel structural model complexity metric, which allows assigning custom symbol complexities in order to incentivize the usage of certain symbols. The model complexity metric is set as an additional objective function, so that candidate models of varying fitness and complexity can be compared after the training.

The gradient-informed GEP framework is applied to four use cases to rediscover Sutherland's law (see Section \ref{sec:sutherland}), develop laminar flame speed models (see Section \ref{sec:combustion}) and train two types of turbulence models for LES (see Section \ref{sec:les}) and RANS calculations (see Section \ref{sec:rans}). All use cases demonstrate significant improvements in prediction accuracy and training convergence speed for more and less complex optimization problems, respectively. While the BFGS optimizer provides more flexibility for defining the objective function, the proof-of-concept use case of rediscovering Sutherland's law already demonstrates the advantages of the LM optimizer for least squares objective functions. However, developing NLEVM for RANS calculations shows that the LM optimizer can converge to extraordinarily high numerical constants without regularization. $L_2$ regularization is effective at maintaining small magnitude constants and even speeds up training convergence for developing SGS models for LES. The model complexity objective function reduces the expression length of the developed NLEVM and allows identifying the optimal trade-off between prediction accuracy and model complexity. Furthermore, setting custom symbol complexity values is shown to be useful to prevent an excessive usage of nonlinear mathematical operators when developing laminar flame speed models. Finally, the implementability of the developed turbulence models, enabled by the introduced regularization methods, is demonstrated by running LES and RANS calculations, which yield promising predictions that outperform state-of-the-art turbulence models.

We consider the concept of adaptive symbols and the two regularization techniques important extensions to the GEP framework to progress towards developing more accurate and generalized closure models while ensuring their implementability. Future research will focus on utilizing adaptive symbols not only in training on high-fidelty data but also in simulation-driven training, which invokes the numerical solver to evaluate candidate models. This approach ensures consistency between the training and prediction environments of the developed models, which is a frequent issue in the practial application of data-driven closure models \cite{duraisamy2021}. However, simulation-driven training is generally expensive and gradient-based optimization requires even more objective function evaluations than the gradient-free standard training approach. Identifying low-order approximations to the numerical solver could be one path to drastically reduce computational costs and enable the development of consistent models with accurate numerical constants.

\section*{Acknowledgments}

Fabian Waschkowski was supported by a Melbourne Research Scholarship provided by the University of Melbourne. Haochen Li and Yaomin Zhao thank the National Natural Science Foundation of China (Grant No. 92152102) for financial support. Richard D. Sandberg acknowledges financial support from the Australian Research Council.

\appendix
\section{Hyperparameter studies for NLEVM use case}
\label{sec:hyperparameters}

\begin{figure}[!ht]
    \centering
    \includegraphics[width=1.0\textwidth]{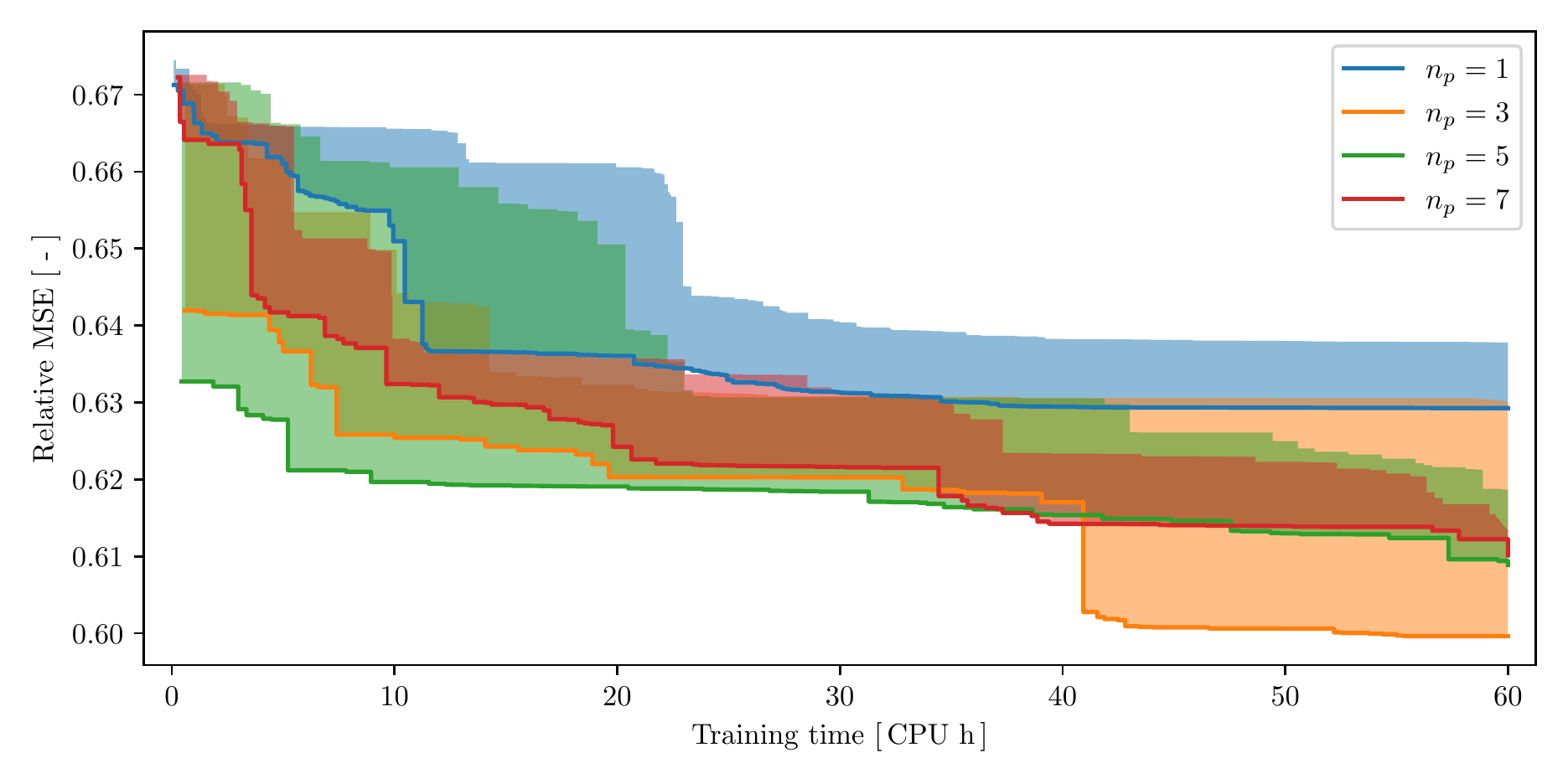}
    \caption{Convergence of MSE relative to LEVM of $R$ models trained with varying number of adaptive symbols per scalar function for NLEVM case (shaded areas represent variation due to random initialization).}
    \label{fig:nlevmadaptive}
\end{figure}

Two hyperparameter studies are performed to determine the optimal number of adaptive symbols per scalar function $n_p$ and the $L_2$ regularization parameter $\lambda$ for the NLEVM use case in Section \ref{sec:rans}. We focus on the development of $R$ models, which present a higher potential for improvement compared to $a_{ij}$ models (see Fig. \ref{fig:nlevmconvergence}), and limit the maximum training runtime to 60 CPU hours in order to reduce the computational costs. The other modeling and training settings are unchanged from Table \ref{tab:strategies}.

To identify the optimal $n_p$ value, we employ the BFGS optimizer, which is expected to require more adaptive symbols than the LM optimizer due to the performance disadvantages observed in Sections \ref{sec:sutherland} and \ref{sec:les}. Fig. \ref{fig:nlevmadaptive} shows the convergence of the MSE relative to the baseline LEVM, which applies no turbulence production correction, for training with $n_p \in \{1, 3, 5, 7\}$ adaptive symbols per scalar function. While $n_p = 1$ is clearly not sufficient to reduce the error to the level of the other approaches, utilizing three adaptive symbols per scalar function achieves the minimum MSE value. However, the orange shaded area indicates a high dependency on the random initialization for the $n_p = 3$ training runs. This dependency is significantly reduced for $n_p = 5$ and $n_p = 7$. Since $n_p = 7$ does not further improve the MSE value compared to $n_p = 5$, we select five adaptive symbols per scalar function as the optimal trade-off between training performance and reliability.

\begin{table}[ht]
	\centering
	\caption{Training and testing MSE relative to LEVM, model complexity $J_c$ and maximum absolute adaptive symbol value $\max_i(|p_i|)$ of $R$ models trained with varying $L_2$ regularization parameter $\lambda$ for NLEVM case.}
    \begin{tabular}{l c c c c}
        \toprule
        $\lambda$ & Train. MSE & Test. MSE & $J_c$ & $\max_i(|p_i|)$ \\
        \cmidrule(lr){1-5}
        $0.0$     & 0.5644 & 0.5558 & 186 & $8.55 \times 10^{17}$ \\
        $10^{-9}$ & 0.5790 & 0.5700 & 170 & $8.72 \times 10^1$ \\
        $10^{-7}$ & 0.5874 & 0.5770 & 164 & $1.12 \times 10^1$ \\
        $10^{-5}$ & 0.6166 & 0.6052 & 194 & $6.97 \times 10^0$ \\
        $10^{-3}$ & 0.6162 & 0.6048 & 226 & $2.18 \times 10^{-1}$ \\
        \bottomrule
    \end{tabular}
	\label{tab:nlevmL2study}
\end{table}

For the $L_2$ regularization parameter study, the LM optimizer is selected, as unregularized NLEVM training with this optimizer leads to very high adaptive symbol values (see Table \ref{tab:nlevmunregularized}). We explore $\lambda \in \{0.0, 10^{-9}, 10^{-7}, 10^{-5}, 10^{-3}\}$ and Table \ref{tab:nlevmL2study} lists the training and testing MSE values, the model complexities $J_c$ and the maximum adaptive symbol magnitudes $\max_i(|p_i|)$ of the resulting $R$ models. Interestingly, a small amount of $L_2$ regularization, realized by setting $\lambda = 10^{-9}$, is sufficient to yield $p$ values below $10^2$. Increasing the regularization parameter to $\lambda = 10^{-7}$ results in a similar testing error and reduces the $\max_i(|p_i|)$ value by nearly one order of magnitude. While the even larger values of $\lambda = 10^{-5}$ and $\lambda = 10^{-3}$ reduce the $\max_i(|p_i|)$ value further, both MSE values and the model complexity increase. Thus, we choose $\lambda = 10^{-7}$ for the regularized training in Section \ref{sec:rans}.

\section{Modeling and training strategies}
\label{sec:strategies}

\begin{sidewaystable}[h]
	\centering
	\caption{Overview of modeling and training strategies for investigated use cases.}
    \begin{tabular}{l c c c c c}
        \toprule
        Use case                                    & Sutherland's law                                      & Laminar flame speed                               & SGS modeling                  & NLEVM modeling \\
        \midrule
        Models                                      & $\mu$, $\hat{\mu}$                                    & $\hat{S}_{L}$                                     & $\tau_{ij}^{A,\text{GEP}}$    & $a_{ij}$/$a_{ij}^R$ \\
        Training objective                          & MSE                                                   & Normalized MSE                                    & MSE                           & MSE \\
        Input symbols                               & $T$                                                   & $\hat{p}$, $\hat{T}$, $\hat{\phi}$                & $I^1, \ldots ,I^4$            & $I^1, \ldots ,I^5$ \\
        Mathematical operators                      & $+$, $-$, $\times$, $\div$, $(\sdot)^\frac{3}{2}$     & $+$, $-$, $\times$, $\exp$, $\log$, $\sqrt{\sdot}$ & $+$, $-$, $\times$           & $+$, $-$, $\times$ \\
        Numerical constants                         & $0$, $1$, $2$                                         & $-1$, $0$, $1$, $2$                               & $-1$, $0$, $1$, $2$           & $-1$, $0$, $1$, $2$ \\
        \# RNCs                                     & $5$/$0$                                               & $5$/$0$                                           & $5$/$0$                       & $5$/$0$ \\
        \# Adaptive symbols                         & $0$/$5$                                               & $0$/$10$                                          & $0$/$16$                      & $0$/$50$ \\
        Optimizers                                  & --/BFGS/LM                                            & --/BFGS                                           & --/BFGS/LM                    & --/BFGS/LM \\
        Regularization                              & --                                                    & --/$J_c$                                          & --/$\lambda = 10^{-5}$        & --/$\lambda = 10^{-7}$, $J_c$ \\
        Population size                             & 1000                                                  & 200                                               & 100                           & 100 \\
        \# Genes                                    & 2                                                     & 5                                                 & 4                             & 10 \\
        \# Initializations                          & 5                                                     & 3                                                 & 5                             & 3 \\
        Max. runtime [$\text{CPU} \, \si{\hour}$]   & 0.25                                                  & 72/120                                            & 8.3                           & 240 \\
        \bottomrule
    \end{tabular}
	\label{tab:strategies}
\end{sidewaystable}

\clearpage
\bibliography{AC_paper_2022}

\end{document}